\documentclass[a4paper,12pt]{article}
\usepackage{amssymb,graphicx,color,a4wide,placeins}
\usepackage[intlimits,fleqn]{amsmath}
\numberwithin{equation}{section}
\newcommand{\p}{\widetilde{p}}

\newcommand{\Li}[1]{\mathop{\mathrm{Li}}\nolimits_{#1}}
\newcommand{\1}[1]{\mathbf{1}^#1}
\newcommand{\2}[1]{\mathbf{2}^#1}
\newcommand{\3}[1]{\mathbf{3}^#1}
\newcommand{\4}[1]{\mathbf{4}^#1}
\newcommand{\5}[1]{\mathbf{5}^#1}

\begin{document}

\begin{flushright}
\Large TTP03-19
\end{flushright}
\vspace{10mm}
\begin{center}
\Huge Lectures on multiloop calculations\\[10mm]
\Large Andrey Grozin\\
Institut f\"ur Theoretische Teilchenphysik,
Universit\"at Karlsruhe\\
grozin@particle.uni-karlsruhe.de
\end{center}

\begin{abstract}
I discuss methods of calculation of propagator diagrams
(massless, those of Heavy Quark Effective Theory, and massive on-shell diagrams)
up to 3 loops.
Integration-by-parts recurrence relations are used to reduce them
to linear combinations of basis integrals.
Non-trivial basis integrals have to be calculated by some other method,
e.g., using Gegenbauer polynomial technique.
Many of them are expressed via hypergeometric functions;
in the massless and HQET cases, their indices tend to integers
at $\varepsilon\to0$.
I discuss the algorithm of their expansion in $\varepsilon$,
in terms of multiple $\zeta$ values.
These lectures were given at Calc-03 school, Dubna, 14--20 June 2003.
\end{abstract}

\newpage
\tableofcontents
\newpage

\section{Introduction}
\label{S:Intro}

We shall discuss methods of calculation of propagator diagrams
in various theories and kinematics, up to 3 loops.
In order to obtain renormalization-group functions,
it is often enough to calculate propagator diagrams.
Massless diagrams are discussed in Sect.~\ref{S:Q};
they are useful, e.g., in QCD with light quarks.
Heavy Quark Effective Theory (HQET) is discussed in Sect.~\ref{S:HQET},
and massive on-shell diagrams -- in Sect.~\ref{S:M}.
Finally, some mathematical methods, most useful in massless
and HQET calculations, are considered in Sect.~\ref{S:hyper}.

Throughout these lectures, we shall discuss scalar Feynman integrals.
Tensor integrals can be expanded in tensor structures,
scalar coefficients are found by solving linear systems.
Their solution can be written as projectors (tensor, $\gamma$-matrix)
applied to the original diagram.

Scalar products of momenta in the numerator can be expressed via the denominators.
In some classes of problems, there are not enough independent denominators
to express all scalar products.
Then we add the required number of scalar products to the basis,
and consider scalar integrals with powers of these scalar products
in the numerators.

\begin{figure}[ht]
\begin{center}
\begin{picture}(94,24)
\put(21,12){\makebox(0,0){\includegraphics{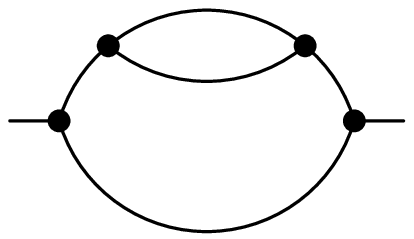}}}
\put(73,12){\makebox(0,0){\includegraphics{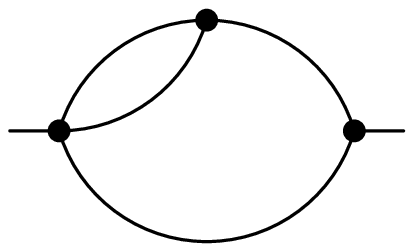}}}
\put(47,12){\makebox(0,0){{${}={}$}}}
\put(6,16){\makebox(0,0)[r]{{$n_1$}}}
\put(36,16){\makebox(0,0)[l]{{$n_2$}}}
\put(86,19){\makebox(0,0)[l]{{$n_1+n_2$}}}
\end{picture}
\end{center}
\caption{Insertion into a propagator}
\label{F:insert}
\end{figure}

Some diagrams contain insertions into internal lines.
The denominators of the propagators on both sides coincide.
If powers of these denominators are $n_1$ and $n_2$,
we can combine them into a single line with the power $n_1+n_2$
(Fig.~\ref{F:insert}).

Throughout these lectures, Minkowski notations with the metric
${}+{}-{}-{}-{}$ are used.

\FloatBarrier
\section{Massless propagator diagrams}
\label{S:Q}

\subsection{1 loop}
\label{S:Q1}

\begin{figure}[ht]
\begin{center}
\begin{picture}(64,42)
\put(32,21){\makebox(0,0){\includegraphics{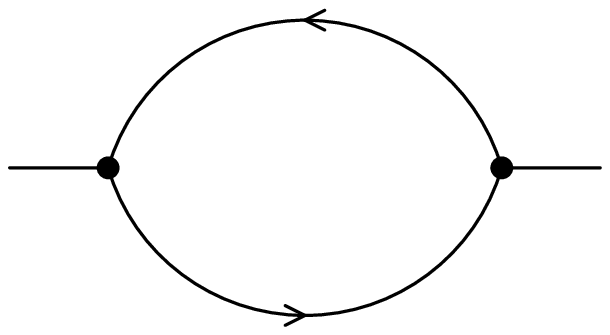}}}
\put(32,0){\makebox(0,0)[b]{{$k+p$}}}
\put(32,42){\makebox(0,0)[t]{{$k$}}}
\put(32,8){\makebox(0,0)[b]{{$n_1$}}}
\put(32,34){\makebox(0,0)[t]{{$n_2$}}}
\end{picture}
\end{center}
\caption{1-loop massless propagator diagram}
\label{F:Q1}
\end{figure}

The 1-loop massless propagator diagram (Fig.~\ref{F:Q1}) is
\begin{equation}
\begin{split}
&\int \frac{d^d k}{D_1^{n_1}D_2^{n_2}} =
i \pi^{d/2} (-p^2)^{d/2-n_1-n_2} G(n_1,n_2)\,,\\
&D_1 = -(k+p)^2-i0\,,\quad
D_2 = -k^2-i0\,.
\end{split}
\label{Q1:G}
\end{equation}
It is symmetric with respect to $1\leftrightarrow 2$.

\begin{figure}[ht]
\begin{center}
\begin{picture}(14,11)
\put(7,5.5){\makebox(0,0){\includegraphics{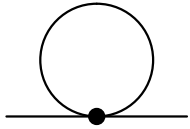}}}
\end{picture}
\end{center}
\caption{1-loop massless vacuum diagram}
\label{F:Q10}
\end{figure}

If $n_1$ is integer and $n_1\le0$,
it becomes the vacuum massless diagram (Fig.~\ref{F:Q10})
with the polynomial $D_1^{|n_1|}$ in the numerator.
It reduces to a sum of powers of $-p^2$ multiplied by integrals
\begin{equation}
\int \frac{d^d k}{(-k^2-i0)^n} = 0\,.
\label{Q1:ScaleFree}
\end{equation}
This integral has the mass dimension $d-2n$;
it contains no dimensionful parameters,
and the only possible result for it is $0$%
\footnote{This argument fails at $n=d/2$;
more careful investigation~\cite{GI:85} shows that
the right-hand side contains $\delta(d-2n)$.}.
Therefore, $G(n_1,n_2)=0$ for non-positive integer $n_1$ (or $n_2$).
More generally, any massless vacuum diagram with any number of loops
is scale-free and hence vanishes.

This diagram is calculated by the Fourier transform to $x$-space and back:
\begin{gather}
\int \frac{e^{-i p\cdot x}}{(-p^2-i0)^n} \frac{d^d p}{(2\pi)^d} =
i 2^{-2n} \pi^{-d/2} \frac{\Gamma(d/2-n)}{\Gamma(n)} \frac{1}{(-x^2+i0)^{d/2-n}}\,,
\label{Q1:p2x}\\
\int \frac{e^{i p\cdot x}}{(-x^2+i0)^n} d^d x =
-i 2^{d-2n} \pi^{d/2} \frac{\Gamma(d/2-n)}{\Gamma(n)} \frac{1}{(-p^2-i0)^{d/2-n}}\,.
\label{Q1:x2p}
\end{gather}
Our diagram in $x$-space is the product of two propagators~(\ref{Q1:p2x})
with the powers $n_1$ and $n_2$:
\begin{equation*}
- 2^{-2(n_1+n_2)} \pi^{-d}
\frac{\Gamma(d/2-n_1) \Gamma(d/2-n_2)}{\Gamma(n_1) \Gamma(n_2)}
\frac{1}{(-x^2)^{d-n_1-n_2}}\,.
\end{equation*}
Transforming it back to $p$-space~(\ref{Q1:x2p}),
we arrive at
\begin{equation}
G(n_1,n_2) =
\frac{\Gamma(-d/2+n_1+n_2) \Gamma(d/2-n_1) \Gamma(d/2-n_2)}%
{\Gamma(n_1) \Gamma(n_2) \Gamma(d-n_1-n_2)}\,.
\label{Q1:G1}
\end{equation}

\begin{figure}[ht]
\begin{center}
\begin{picture}(74,26)
\put(16,13){\makebox(0,0){\includegraphics{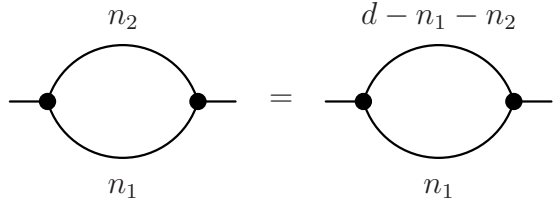}}}
\put(58,13){\makebox(0,0){\includegraphics{q1.eps}}}
\put(37,13){\makebox(0,0){{${}={}$}}}
\put(16,0){\makebox(0,0)[b]{{$n_1$}}}
\put(16,26){\makebox(0,0)[t]{{$\vphantom{d}n_2$}}}
\put(58,0){\makebox(0,0)[b]{{$n_1$}}}
\put(58,26){\makebox(0,0)[t]{{$d-n_1-n_2$}}}
\end{picture}
\end{center}
\caption{Inversion relation}
\label{F:Q1i}
\end{figure}

Diagrams are calculated by going to Euclidean momentum space:
$k_0=ik_{E0}$, $k^2=-k_E^2$.
Using a dimensionless loop momentum $K=k_E/\sqrt{-p^2}$,
we can rewrite~(\ref{Q1:G}) as
\begin{equation*}
\int \frac{d^d K}{\left[(K+n)^2\right]^{n_1} \left[K^2\right]^{n_2}}
= \pi^{d/2} G(n_1,n_2)\,,
\end{equation*}
where $n$ is a unit Euclidean vector ($n^2=1$).
We can perform inversion
\begin{equation}
K = \frac{K'}{K^{\prime2}}\,,\quad
K^2 = \frac{1}{K^{\prime2}}\,,\quad
d^d K = \frac{d^d K'}{(K^{\prime2})^d}\,.
\label{Q1:Inv}
\end{equation}
Then
\begin{equation*}
(K+n)^2 = \frac{(K'+n)^2}{K^{\prime2}}\,,
\end{equation*}
and we obtain (Fig.~\ref{F:Q1i})
\begin{equation}
G(n_1,n_2) = G(n_1,d-n_1-n_2)\,.
\label{Q1:i}
\end{equation}

\subsection{2 loops}
\label{S:Q2}

\begin{figure}[ht]
\begin{center}
\begin{picture}(64,32)
\put(32,18){\makebox(0,0){\includegraphics{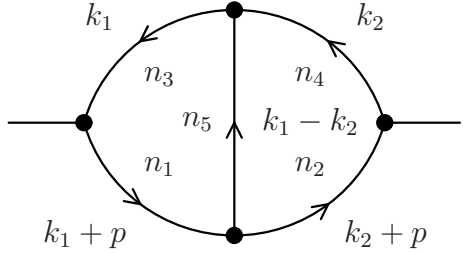}}}
\put(14,32){\makebox(0,0){{$k_1$}}}
\put(50,32){\makebox(0,0){{$k_2$}}}
\put(12,3){\makebox(0,0){{$k_1+p$}}}
\put(52,3){\makebox(0,0){{$k_2+p$}}}
\put(42,18){\makebox(0,0){{$k_1-k_2$}}}
\put(22,12){\makebox(0,0){{$n_1$}}}
\put(42,12){\makebox(0,0){{$n_2$}}}
\put(22,24){\makebox(0,0){{$n_3$}}}
\put(42,24){\makebox(0,0){{$n_4$}}}
\put(27,18){\makebox(0,0){{$n_5$}}}
\end{picture}
\end{center}
\caption{2-loop massless propagator diagram}
\label{F:Q2}
\end{figure}

There is 1 generic topology of 2-loop massless propagator diagrams
(Fig.~\ref{F:Q2}):
\begin{gather}
\int \frac{d^d k_1\,d^d k_2}{D_1^{n_1} D_2^{n_2} D_3^{n_3} D_4^{n_4} D_5^{n_5}} =
- \pi^d (-p^2)^{d-\sum n_i} G(n_1,n_2,n_3,n_4,n_5)\,,
\label{Q2:G}\\
D_1=-(k_1+p)^2\,,\quad
D_2=-(k_2+p)^2\,,\quad
D_3=-k_1^2\,,\quad
D_4=-k_2^2\,,\quad
D_5=-(k_1-k_2)^2\,.
\nonumber
\end{gather}
All other diagrams (e.g., Fig.~\ref{F:insert})
are particular cases of this one,
when some line shrinks to a point,
i.e., its index $n_i=0$.
This diagram is symmetric with respect to $(1\leftrightarrow2,3\leftrightarrow4)$,
and with respect to $(1\leftrightarrow3,2\leftrightarrow4)$.
If indices of any two adjacent lines are non-positive,
the diagram contains a scale-free vacuum subdiagram,
and hence vanishes.

\begin{figure}[ht]
\begin{center}
\begin{picture}(52,26)
\put(26,13){\makebox(0,0){\includegraphics{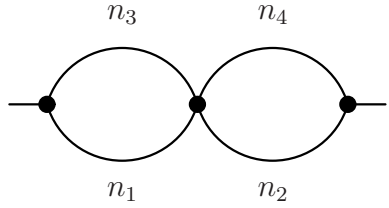}}}
\put(16,0){\makebox(0,0)[b]{{$n_1$}}}
\put(36,0){\makebox(0,0)[b]{{$n_2$}}}
\put(16,26){\makebox(0,0)[t]{{$n_3$}}}
\put(36,26){\makebox(0,0)[t]{{$n_4$}}}
\end{picture}
\end{center}
\caption{Trivial case $n_5=0$}
\label{F:Q2z5}
\end{figure}

\begin{figure}[ht]
\begin{center}
\begin{picture}(134,32)
\put(25,17.5){\makebox(0,0){\includegraphics{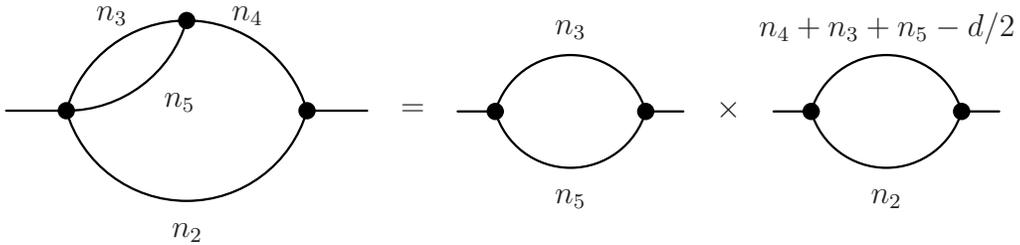}}}
\put(76,17.5){\makebox(0,0){\includegraphics{q1.eps}}}
\put(118,17.5){\makebox(0,0){\includegraphics{q1.eps}}}
\put(55,17.5){\makebox(0,0){{${}={}$}}}
\put(97,17.5){\makebox(0,0){{${}\times{}$}}}
\put(25,0){\makebox(0,0)[b]{{$n_2$}}}
\put(15,30){\makebox(0,0){{$n_3$}}}
\put(33,30){\makebox(0,0){{$n_4$}}}
\put(24,18.5){\makebox(0,0){{$n_5$}}}
\put(76,4.5){\makebox(0,0)[b]{{$n_5$}}}
\put(76,30.5){\makebox(0,0)[t]{{$\vphantom{d}n_3$}}}
\put(118,4.5){\makebox(0,0)[b]{{$n_2$}}}
\put(118,30.5){\makebox(0,0)[t]{{$n_4+n_3+n_5-d/2$}}}
\end{picture}
\end{center}
\caption{Trivial case $n_1=0$}
\label{F:Q2z1}
\end{figure}

If $n_5=0$, our diagram is just the product of two 1-loop diagrams
(Fig.~\ref{F:Q2z5}):
\begin{equation}
G(n_1,n_2,n_3,n_4,0) = G(n_1,n_3) G(n_2,n_4)\,.
\label{Q2:z5}
\end{equation}
If $n_1=0$, the inner loop gives $G(n_3,n_5)/(-k_2^2)^{n_3+n_5-d/2}$,
and (Fig.~\ref{F:Q2z1})
\begin{equation}
G(0,n_2,n_3,n_4,n_5) = G(n_3,n_5) G(n_2,n_4+n_3+n_5-d/2)\,.
\label{Q2:z1}
\end{equation}
The cases $n_2=0$, $n_3=0$, $n_4=0$ are symmetric.

When all $n_i>0$, the problem does not immediately reduce
to a repeated use of the 1-loop formula~(\ref{Q1:G1}).
We shall use a powerful method
called integration by parts~\cite{CT:81}.
It is based on the simple observation that any integral of
$\partial/\partial k_1(\cdots)$
(or $\partial/\partial k_2(\cdots)$)
vanishes (in dimensional regularization no surface terms can appear).
From this, we can obtain recurrence relations which involve
$G(n_1,n_2,n_3,n_4,n_5)$ with different sets of indices.
Applying these relations in a carefully chosen order,
we can reduce any $G(n_1,n_2,n_3,n_4,n_5)$ to trivial ones,
like~(\ref{Q2:z5}), (\ref{Q2:z1}).

The differential operator $\partial/\partial k_2$
applied to the integrand of~(\ref{Q2:G}) acts as
\begin{equation}
\frac{\partial}{\partial k_2} \to
\frac{n_2}{D_2}2(k_2+p) + \frac{n_4}{D_4}2k_2 + \frac{n_5}{D_5}2(k_2-k_1)\,.
\label{Q2:dk1}
\end{equation}
Applying $(\partial/\partial k_2)\cdot k_2$ to the integrand of~(\ref{Q2:G}),
we get a vanishing integral.
On the other hand, from~(\ref{Q2:G}), $2k_2\cdot k_2=-2D_4$,
$2(k_2+p)\cdot k_2=(-p^2)-D_2-D_4$, $2(k_2-k_1)\cdot k_2=D_3-D_4-D_5$,
we see that this differential operator is equivalent to inserting
\begin{equation*}
d-n_2-n_5-2n_4 + \frac{n_2}{D_2}((-p^2)-D_4) + \frac{n_5}{D_5}(D_3-D_4)
\end{equation*}
under the integral sign (here $(\partial/\partial k_2)\cdot k_2=d$).
Taking into account the definition~(\ref{Q2:G}),
we obtain the recurrence relation
\begin{equation}
\left[d-n_2-n_5-2n_4 + n_2\2+(1-\4-) + n_5\5+(\3--\4-)\right] G = 0\,.
\label{Q2:Tri1}
\end{equation}
Here
\begin{equation}
\1\pm G(n_1,n_2,n_3,n_4,n_5) = G(n_1\pm1,n_2,n_3,n_4,n_5)\,,
\label{Q2:pm}
\end{equation}
and similar ones for the other indices.

This is a particular example of the triangle relation.
We differentiate in the loop momentum running along the triangle 254,
and insert the momentum of the line 4 in the numerator.
The differentiation raises the degree of one of the denominators 2, 5, 4.
In the case of the line 4, we get $-2D_4$ in the numerator,
giving just $-2n_4$.
In the case of the line 5, we get the denominator $D_3$
of the line attached to the vertex 45 of our triangle,
minus the denominators $D_4$ and $D_5$.
The case of the line 2 is similar; the denominator of the line
attached to the vertex 24 of our triangle is just $-p^2$,
and it does not influence any index of $G$.
Of course, there are three more relations obtained from~(\ref{Q2:Tri1})
by the symmetry.
Another useful triangle relation is derived by applying the operator
$(\partial/\partial k_2)\cdot(k_2-k_1)$:
\begin{equation}
\left[d-n_2-n_4-2n_5 + n_2\2+(\1--\5-) + n_4\4+(\3--\5-)\right] G = 0\,.
\label{Q2:Tri2}
\end{equation}
One more is obtained by the symmetry.
Relations of this kind can be written for any diagram having a triangle in it,
when at least two vertices of the triangle each have only a single line
(not belonging to the triangle) attached.

We can obtain a relation from homogeneity of the integral~(\ref{Q2:G}) in $p$.
Applying the operator $p\cdot(\partial/\partial p)$ to the integral~(\ref{Q2:G}),
we see that it is equivalent to the factor $2(d-\sum n_i)$.
On the other hand, explicit differentiation of the integrand gives
$-(n_1/D_1)(-p^2+D_1-D_3)-(n_2/D_2)(-p^2+D_2-D_4)$.
Therefore,
\begin{equation}
\left[ 2(d-n_3-n_4-n_5)-n_1-n_2 + n_1\1+(1-\3-) + n_2\2+(1-\4-) \right] I = 0\,.
\label{Q2:hom}
\end{equation}
This is nothing but the sum of the $(\partial/\partial k_2)\cdot k_2$
relation~(\ref{Q2:Tri1}) and its mirror-symmetric
$(\partial/\partial k_1)\cdot k_1$ relation.

Another interesting relation is obtained by inserting $(k_1+p)^\mu$
into the integrand of~(\ref{Q2:G}) and taking derivative $\partial/\partial p^\mu$
of the integral.
On the one hand, the vector integral must be proportional to $p^\mu$,
and we can make the substitution
\begin{equation*}
k_1+p \to \frac{(k_1+p)\cdot p}{p^2} p =
\left(1 + \frac{D_1-D_3}{-p^2}\right) \frac{p}{2}
\end{equation*}
in the integrand.
Taking $\partial/\partial p^\mu$ of this vector integral produces~(\ref{Q2:G}) with
\begin{equation*}
\left(\tfrac{3}{2}d-\sum n_i\right)
\left(1 + \frac{D_1-D_3}{-p^2}\right)
\end{equation*}
inserted into the integrand.
On the other hand, explicit differentiation in $p$ gives
\begin{gather*}
d + \frac{n_1}{D_1} 2(k_1+p)^2 + \frac{n_2}{D_2} 2(k_2+p)\cdot(k_1+p)\,,\\
2(k_2+p)\cdot(k_1+p) = D_5-D_1-D_2\,.
\end{gather*}
Therefore, we obtain
\begin{equation}
\left[\tfrac{1}{2}d+n_1-n_3-n_4-n_5
+ \left(\tfrac{3}{2}d-\sum n_i\right)(\1--\3-)
+ n_2\2+(\1--\5-)\right] G = 0\,.
\label{Q2:Larin}
\end{equation}
This relation has been derived by S.A.~Larin in his M.~Sc.\ thesis.
Three more relations follow from the symmetries.

\begin{figure}[p]
\begin{center}
\begin{picture}(82,82)
\put(41,41){\makebox(0,0){\includegraphics{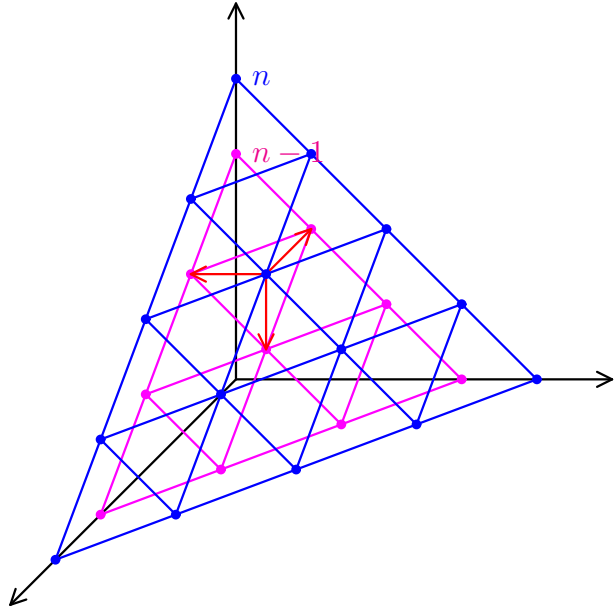}}}
\put(33,71){\makebox(0,0)[l]{\textcolor{blue}{$n$}}}
\put(33,61){\makebox(0,0)[l]{\textcolor{magenta}{$n-1$}}}
\end{picture}
\end{center}
\caption{Reduction of $n=n_1+n_3+n_5$}
\label{F:red}
\end{figure}

Expressing $G(n_1,n_2,n_3,n_4,n_5)$ from~(\ref{Q2:Tri2}):
\begin{equation}
G(n_1,n_2,n_3,n_4,n_5) =
\frac{n_2\2+(\5--\1-) + n_4\4+(\5--\3-)}{d-n_2-n_4-2n_5} G\,,
\label{Q2:Tri}
\end{equation}
we see that the sum $n_1+n_3+n_5$ reduces by 1.
If we start from an integral belonging to the plane $n_1+n_3+n_5=n$
in Fig.~\ref{F:red}, then each of the integrals in the right-hand side
belong to the plane $n_1+n_3+n_5=n-1$ (Fig.~\ref{F:red}).
Therefore, applying~(\ref{Q2:Tri}) sufficiently many times,
we can reduce an arbitrary $G$ integral with integer indices
to a combination of integrals
with $n_5=0$ (Fig.~\ref{F:Q2z5}, (\ref{Q2:z5})),
$n_1=0$ (Fig.~\ref{F:Q2z1}, (\ref{Q2:z1})),
$n_3=0$ (mirror-symmetric to the previous case).
Of course, if $\max(n_2,n_4)<\max(n_1,n_3)$, it may be more efficient
to use the relation mirror-symmetric to~(\ref{Q2:Tri2}).
The relation~(\ref{Q2:Larin}) also can be used instead of~(\ref{Q2:Tri2}).

\begin{figure}[p]
\begin{center}
\begin{picture}(32,17)
\put(16,8.5){\makebox(0,0){\includegraphics{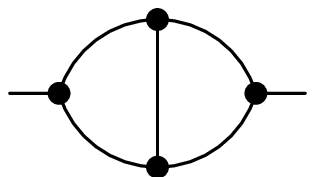}}}
\end{picture}
\end{center}
\caption{2-loop massless propagator diagram}
\label{F:Q2t}
\end{figure}

\begin{figure}[p]
\begin{center}
\begin{picture}(104,17)
\put(11,8.5){\makebox(0,0){\includegraphics{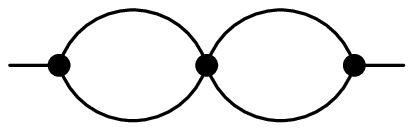}}}
\put(33,8.5){\makebox(0,0)[l]{{${}=G_1^2$}}}
\put(68,8.5){\makebox(0,0){\includegraphics{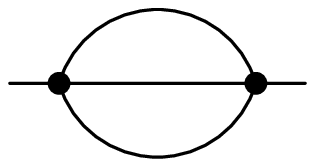}}}
\put(85,8.5){\makebox(0,0)[l]{{${}=G_2$}}}
\end{picture}
\end{center}
\caption{Basis diagrams (all indices equal 1)}
\label{F:Q2b}
\end{figure}

\begin{figure}[p]
\begin{center}
\begin{picture}(44,26)
\put(22,13){\makebox(0,0){\includegraphics{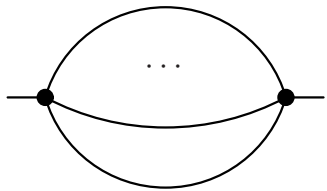}}}
\put(22,17){\makebox(0,0){{$\cdots$}}}
\end{picture}
\end{center}
\caption{$n$-loop massless sunset diagrams}
\label{F:Qsun}
\end{figure}

Let's summarize.
There is 1 generic topology of 2-loop massless propagator diagrams
(Fig.~\ref{F:Q2t}).
The integrals with all integer $n_i$
can be expressed as linear combinations of 2 basis integrals
(Fig.~\ref{F:Q2b});
coefficients are rational functions of $d$.
Here $G_n$ are $n$-loop massless sunset diagrams (Fig.~\ref{F:Qsun}):
\begin{equation}
G_n = \frac{1}{\left(n+1-n\frac{d}{2}\right)_n
\left((n+1)\frac{d}{2}-2n-1\right)_n}
\frac{\Gamma(1+n\varepsilon)\Gamma^{n+1}(1-\varepsilon)}%
{\Gamma(1-(n+1)\varepsilon)}\,.
\label{Q2:sun}
\end{equation}

\begin{figure}[ht]
\begin{center}
\begin{picture}(98,34)
\put(12,17){\makebox(0,0){\includegraphics{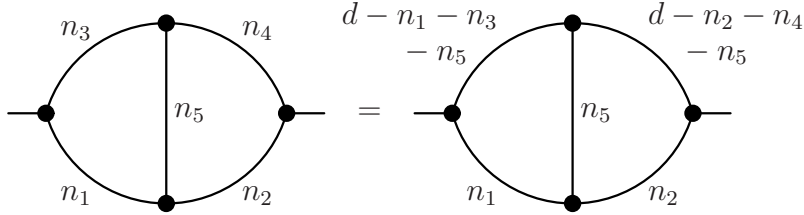}}}
\put(66,17){\makebox(0,0){\includegraphics{q2.eps}}}
\put(39,17){\makebox(0,0){{${}={}$}}}
\put(2,6){\makebox(0,0)[r]{{$n_1$}}}
\put(22,6){\makebox(0,0)[l]{{$n_2$}}}
\put(2,28){\makebox(0,0)[r]{{$n_3$}}}
\put(22,28){\makebox(0,0)[l]{{$n_4$}}}
\put(13,17){\makebox(0,0)[l]{{$n_5$}}}
\put(56,6){\makebox(0,0)[r]{{$n_1$}}}
\put(76,6){\makebox(0,0)[l]{{$n_2$}}}
\put(56,30){\makebox(0,0)[r]{{$d-n_1-n_3$}}}
\put(52,25){\makebox(0,0)[r]{{${}-n_5$}}}
\put(76,30){\makebox(0,0)[l]{{$d-n_2-n_4$}}}
\put(80,25){\makebox(0,0)[l]{{${}-n_5$}}}
\put(67,17){\makebox(0,0)[l]{{$n_5$}}}
\end{picture}
\end{center}
\caption{Inversion relation}
\label{F:Q2i}
\end{figure}

Using inversion~(\ref{Q1:Inv}) of both loop momenta and
\begin{equation*}
(K_1-K_2)^2 = \frac{(K_1'-K_2')^2}{K_1^{\prime2} K_2^{\prime2}}\,,
\end{equation*}
we obtain relation of Fig.~\ref{F:Q2i}%
\footnote{This is one of the elements of the symmetry group $Z_2\times S_6$~\cite{B:86}
(with 1440 elements) of the diagram of Fig.~\ref{F:Q2t}.
If we connect the external vertices by a line having $n_6=\frac{3}{2}d-\sum_{i=1}^5 n_i$,
we obtain a logarithmically divergent 3-loop tetrahedron vacuum diagram
It is proportional to our original diagram times the logarithm of the ultraviolet cut-off.
We can cut it at any of its lines, and get a diagram of Fig.~\ref{F:Q2t}
with 5 indices $n_i$ out of 6.
This gives the tetrahedron symmetry group.
Also, we have the inversion relations (Fig.~\ref{F:Q2i}).
Fourier transform to $x$-space gives a new integral of the form of Fig.~\ref{F:Q2t}
with $n_i\to\frac{d}{2}-n_i$.
There is an additional symmetry following from
the star-triangle relation~\cite{K:84,K:85}.}.

\subsection{3 loops}
\label{S:Q3}

\begin{figure}[p]
\begin{center}
\begin{picture}(94,60)
\put(47,45){\makebox(0,0){\includegraphics{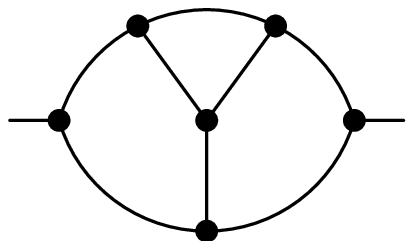}}}
\put(21,15){\makebox(0,0){\includegraphics{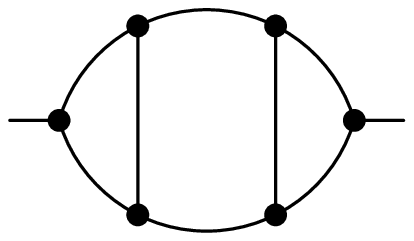}}}
\put(73,15){\makebox(0,0){\includegraphics{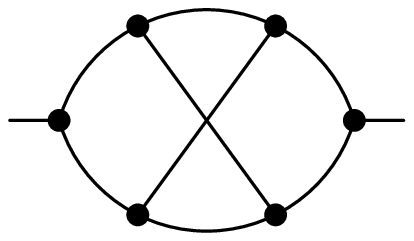}}}
\end{picture}
\end{center}
\caption{Topologies of 3-loop massless propagator diagrams}
\label{F:Q3t}
\end{figure}

\begin{figure}[p]
\begin{center}
\begin{picture}(129,77)
\put(21,68.5){\makebox(0,0){\includegraphics{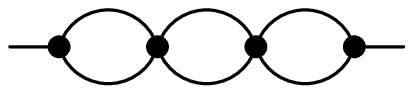}}}
\put(43,68.5){\makebox(0,0)[l]{{${}=G_1^3$}}}
\put(83,68.5){\makebox(0,0){\includegraphics{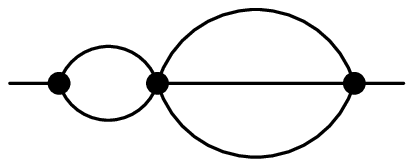}}}
\put(105,68.5){\makebox(0,0)[l]{{${}=G_1 G_2$}}}
\put(21,42.5){\makebox(0,0){\includegraphics{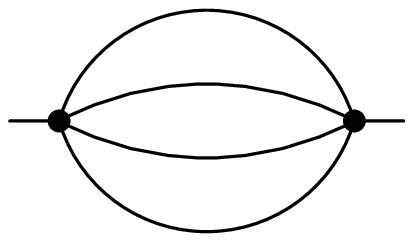}}}
\put(43,42.5){\makebox(0,0)[l]{{${}=G_3$}}}
\put(83,42.5){\makebox(0,0){\includegraphics{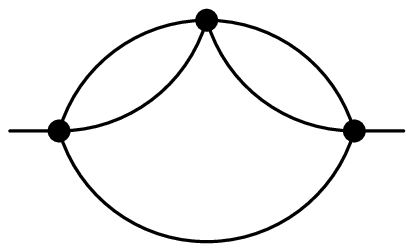}}}
\put(105,42.5){\makebox(0,0)[l]{{$\displaystyle{}\sim G_3\frac{G_1^2}{G_2}$}}}
\put(21,12.5){\makebox(0,0){\includegraphics{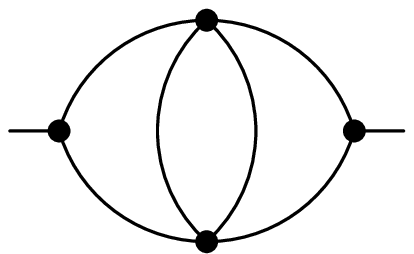}}}
\put(43,12.5){\makebox(0,0)[l]{{${}=G_1 G(1,1,1,1,\varepsilon)$}}}
\put(108,12.5){\makebox(0,0){\includegraphics{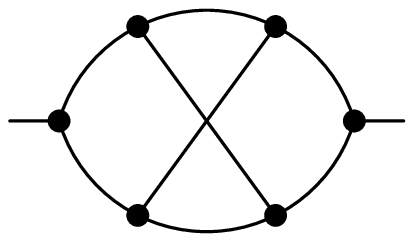}}}
\end{picture}
\end{center}
\caption{Basis diagrams (all indices equal 1, no numerators)}
\label{F:Q3b}
\end{figure}

There are 3 generic topologies of 3-loop massless propagator diagrams
(Fig.~\ref{F:Q3t}).
Each has 8 denominators.
There are 9 scalar products of 3 loop momenta $k_i$ and the external momentum $p$.
Therefore, for each topology, all scalar products in the numerator
can be expressed via the denominators and one selected scalar product.
Integration-by-parts recurrence relations for these diagrams
have been investigated in~\cite{CT:81}.
They can be used to reduce all integrals of Fig.~\ref{F:Q3t},
with arbitrary integer powers of denominators
and arbitrary (non-negative) powers of the selected scalar product
in the numerators,
to linear combinations of 6 basis integrals (Fig.~\ref{F:Q3b}).
This algorithm has been implemented in the \textsf{SCHOONSCHIP}~\cite{SCH}
package \textsf{MINCER}~\cite{MIN} and later re-implemented~\cite{MIN2}
in \textsf{FORM}~\cite{FORM}.
It has also been implemented in the \textsf{REDUCE}~\cite{RED,G:97}
package \textsf{Slicer}~\cite{B:92a}
Only the last, non-planar, topology in Fig.~\ref{F:Q3t}
involve the last, non-planar, basis integral in Fig.~\ref{F:Q3b}.

\begin{figure}[p]
\begin{center}
\begin{picture}(102,70)
\put(11,52.5){\makebox(0,0){\includegraphics{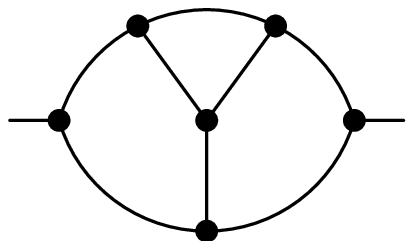}}}
\put(73,52.5){\makebox(0,0){\includegraphics{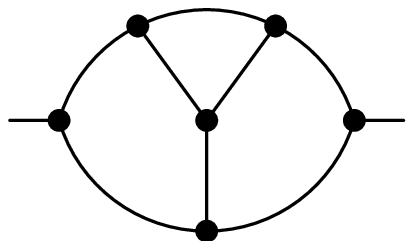}}}
\put(37,52.5){\makebox(0,0){{${}={}$}}}
\put(11,17.5){\makebox(0,0){\includegraphics{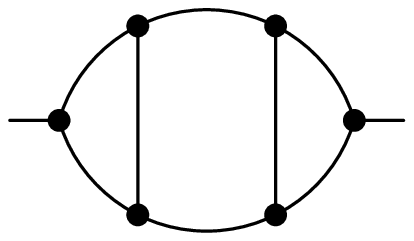}}}
\put(73,17.5){\makebox(0,0){\includegraphics{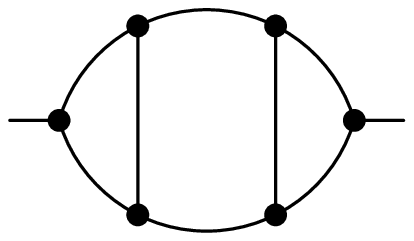}}}
\put(37,17.5){\makebox(0,0){{${}={}$}}}
\put(1,43){\makebox(0,0)[r]{{$n_1$}}}
\put(21,43){\makebox(0,0)[l]{{$n_2$}}}
\put(-2,59){\makebox(0,0)[r]{{$n_3$}}}
\put(24,59){\makebox(0,0)[l]{{$n_4$}}}
\put(12,46){\makebox(0,0)[l]{{$n_5$}}}
\put(11,65){\makebox(0,0)[b]{{$n_6$}}}
\put(7.5,55){\makebox(0,0)[r]{{$n_7$}}}
\put(14.5,55){\makebox(0,0)[l]{{$n_8$}}}
\put(63,43){\makebox(0,0)[r]{{$n_1$}}}
\put(83,43){\makebox(0,0)[l]{{$n_2$}}}
\put(62,62){\makebox(0,0)[r]{{$d-n_1-n_3$}}}
\put(59,58){\makebox(0,0)[r]{{${}-n_5-n_7$}}}
\put(84,62){\makebox(0,0)[l]{{$d-n_2-n_4$}}}
\put(87,58){\makebox(0,0)[l]{{${}-n_5-n_8$}}}
\put(74,46){\makebox(0,0)[l]{{$n_5$}}}
\put(73,65){\makebox(0,0)[b]{{$d-n_6-n_7-n_8$}}}
\put(69.5,55){\makebox(0,0)[r]{{$n_7$}}}
\put(76.5,55){\makebox(0,0)[l]{{$n_8$}}}
\put(-2,11){\makebox(0,0)[r]{{$n_1$}}}
\put(24,11){\makebox(0,0)[l]{{$n_2$}}}
\put(11,5){\makebox(0,0)[t]{{$n_3$}}}
\put(-2,24){\makebox(0,0)[r]{{$n_4$}}}
\put(24,24){\makebox(0,0)[l]{{$n_5$}}}
\put(5,17.5){\makebox(0,0)[l]{{$n_6$}}}
\put(17,17.5){\makebox(0,0)[r]{{$n_7$}}}
\put(11,30){\makebox(0,0)[b]{{$n_8$}}}
\put(60,11){\makebox(0,0)[r]{{$n_1$}}}
\put(86,11){\makebox(0,0)[l]{{$n_2$}}}
\put(73,5){\makebox(0,0)[t]{{$n_3$}}}
\put(62,27){\makebox(0,0)[r]{{$d-n_1-n_3$}}}
\put(59,23){\makebox(0,0)[r]{{${}-n_6$}}}
\put(84,27){\makebox(0,0)[l]{{$d-n_2-n_4$}}}
\put(87,23){\makebox(0,0)[l]{{${}-n_7$}}}
\put(67,17.5){\makebox(0,0)[l]{{$n_6$}}}
\put(79,17.5){\makebox(0,0)[r]{{$n_7$}}}
\put(73,30){\makebox(0,0)[b]{{$d-n_3-n_6-n_7-n_8$}}}
\end{picture}
\end{center}
\caption{Inversion relations}
\label{F:Q3i}
\end{figure}

\begin{figure}[p]
\begin{center}
\begin{picture}(94,25)
\put(21,12.5){\makebox(0,0){\includegraphics{q3t2.eps}}}
\put(73,12.5){\makebox(0,0){\includegraphics{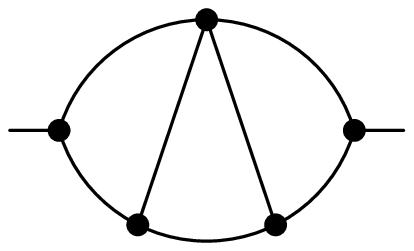}}}
\put(47,12.5){\makebox(0,0){${}={}$}}
\end{picture}
\end{center}
\caption{Inversion relation: all $n_i=1$, $d=4$}
\label{F:Q3i2}
\end{figure}

The first 4 basis integrals are trivial: they are expressed via $G_n$~(\ref{Q2:sun})
(Fig.~\ref{F:Qsun}), and hence via $\Gamma$ functions.
The 4-th one differs from the 3-rd one ($G_3$)
by replacing the 2-loop subdiagram:
the second one in Fig.~\ref{F:Q2b} ($G_2/(-k^2)^{3-d}$)
by the first one ($G_1^2/(-k^2)^{4-d}$).
Therefore, it can be obtained from $G_3$
by multiplying by
\begin{equation*}
\frac{G_1^2 G(1,4-d)}{G_2 G(1,3-d)} = \frac{2d-5}{d-3} \frac{G_1^2}{G_2}\,.
\end{equation*}
The 5-th basis integral is proportional to the 2-loop diagram
$G(1,1,1,1,n)$ with a non-integer index of the middle line $n=\varepsilon$,
and will be discussed in Sect.~\ref{S:Gn}.
The 6-th one, non-planar, is truly 3-loop and most difficult;
it will be discussed in Sect.~\ref{S:NP}.

Performing inversion~(\ref{Q1:Inv}) of the loop momenta,
we obtain the relations in Fig.~\ref{F:Q3i}.
For example, the ladder diagram with all indices $n_i=1$ is convergent;
its value at $d=4$ is related to a simpler diagram (Fig.~\ref{F:Q3i2})
by the second inversion relation.
The non-planar topology (Fig.~\ref{F:Q3t})
involves lines with sums of 3 momenta;
they don't transform into anything reasonable under inversion,
and there is no inversion relation for it.

\subsection{$G(1,1,1,1,n)$}
\label{S:Gn}

\begin{figure}[ht]
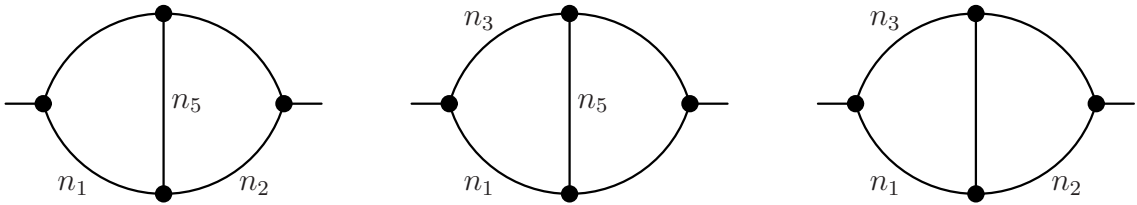

\begin{center}
\begin{picture}(152,26)
\put(22,13){\makebox(0,0){\includegraphics{q2.eps}}}
\put(12,2){\makebox(0,0)[r]{{$n_1$}}}
\put(32,2){\makebox(0,0)[l]{{$n_2$}}}
\put(23,13){\makebox(0,0)[l]{{$n_5$}}}
\put(76,13){\makebox(0,0){\includegraphics{q2.eps}}}
\put(66,2){\makebox(0,0)[r]{{$n_1$}}}
\put(66,24){\makebox(0,0)[r]{{$n_3$}}}
\put(77,13){\makebox(0,0)[l]{{$n_5$}}}
\put(130,13){\makebox(0,0){\includegraphics{q2.eps}}}
\put(120,2){\makebox(0,0)[r]{{$n_1$}}}
\put(140,2){\makebox(0,0)[l]{{$n_2$}}}
\put(120,24){\makebox(0,0)[r]{{$n_3$}}}
\end{picture}
\end{center}
\caption{Diagrams calculated in~\protect\cite{K:96}}
\label{F:Gn}
\end{figure}

This diagram is~\cite{K:96}
\begin{gather}
G(1,1,1,1,n) = 2
\Gamma\left(\tfrac{d}{2}-1\right) \Gamma\left(\tfrac{d}{2}-n-1\right)
\Gamma(n-d+3)\times{}
\label{Gn:F}\\
\left[
\frac{2\Gamma\left(\frac{d}{2}-1\right)}%
{(d-2n-4)\Gamma(n+1)\Gamma\left(\frac{3}{2}d-n-4\right)}
{}_3 F_2 \left(
\begin{array}{c}
1,d-2,n-\frac{d}{2}+2\\n+1,n-\frac{d}{2}+3
\end{array}
\right| \left.\vphantom{\frac{1}{1}} 1\right)
- \frac{\pi\cot\pi(d-n)}{\Gamma(d-2)}
\right]\,.
\nonumber
\end{gather}
This is a particular case of a more general result~\cite{K:96}
for the diagram with 3 non-integer indices.
The two lines with unit indices must be adjacent (Fig.~\ref{F:Gn});
all such diagrams are equivalint, due to the tetrahedron symmetry mentioned above.
This result was derived using Gegenbauer polynomial technique~\cite{CKT:80};
we shall discuss it for a simpler example in Sect.~\ref{S:In}.
When $n$ is integer, the ${}_3 F_2$ in~(\ref{Gn:F}) is expressed via $\Gamma$ functions;
in order to show equivalence to the standard results (Sect.~\ref{S:Q2}),
one should use
\begin{equation*}
\Gamma(1+l\varepsilon) \Gamma(1-l\varepsilon) =
\frac{\pi l\varepsilon}{\sin \pi l\varepsilon}
\end{equation*}
and trigonometric identities.

Similar expressions for these diagrams
were also obtained in~\cite{BGK:97}%
\footnote{The statement that these diagrams can be expressed via
${}_3 F_2$ functions of unit argument was communicated by D.~Broadhurst
to A.~Kotikov before~\cite{K:96}.}.
In our particular case, they reduce to
\begin{equation}
\begin{split}
&\frac{(d-3)(d-4)\Gamma(n)\Gamma\left(\frac{3}{2}d-n-4\right)}%
{2\Gamma(n+3-d)\Gamma^2\left(\frac{d}{2}-1\right)\Gamma\left(\frac{d}{2}-n-1\right)}
G(1,1,1,1,n)\\
&{} = \frac{3d-2n-10}{d-n-3} {}_3 F_2 \left(
\begin{array}{c}1,\frac{d}{2}-2,n+3-d\\n,n+4-d\end{array}\right|
\left.\vphantom{\frac{1}{1}}1\right)\\
&\hphantom{{}={}}{}
+ \frac{\Gamma(n)\Gamma\left(\frac{3}{2}d-n-4\right)}%
{\Gamma(d-4)\Gamma\left(\frac{d}{2}-1\right)}
\pi \cot\pi(n-d)
- 2 \Gamma\left(\tfrac{d}{2}-1\right)\\
&{} = - \frac{3d-2n-10}{d-n-3} {}_3 F_2 \left(
\begin{array}{c}1,1-n,d-n-3\\3-\frac{d}{2},d-n-2\end{array}\right|
\left.\vphantom{\frac{1}{1}}1\right)\\
&\hphantom{{}={}}{}
+ \frac{\Gamma(n)\Gamma\left(\frac{3}{2}d-n-4\right)}%
{\Gamma(d-4)\Gamma\left(\frac{d}{2}-1\right)}
\pi \cot\pi\tfrac{d}{2}
+ \frac{d-4}{d-n-3} \Gamma\left(\tfrac{d}{2}-1\right)\\
&{} = 4 \frac{n-1}{d-2n-2} {}_3 F_2 \left(
\begin{array}{c}1,n+5-\frac{3}{2}d,n+1-\frac{d}{2}\\3-\frac{d}{2},n+2-\frac{d}{2}\end{array}\right|
\left.\vphantom{\frac{1}{1}}1\right)\\
&\hphantom{{}={}}{}
+ \frac{\Gamma(n)\Gamma\left(\frac{3}{2}d-n-4\right)}%
{\Gamma(d-4)\Gamma\left(\frac{d}{2}-1\right)}
\pi \cot\pi\tfrac{d}{2}
- 2 \frac{d-4}{d-2n-2} \Gamma\left(\tfrac{d}{2}-1\right)\\
&{} = - 4 \frac{n-1}{d-2n-2} {}_3 F_2 \left(
\begin{array}{c}1,\frac{d}{2}-2,\frac{d}{2}-n-1\\\frac{3}{2}d-n-4,\frac{d}{2}-n\end{array}\right|
\left.\vphantom{\frac{1}{1}}1\right)\\
&\hphantom{{}={}}{}
+ \frac{\Gamma(n)\Gamma\left(\frac{3}{2}d-n-4\right)}%
{\Gamma(d-4)\Gamma\left(\frac{d}{2}-1\right)}
\pi \cot\pi\left(\tfrac{d}{2}-n\right)
- 2 \Gamma\left(\tfrac{d}{2}-1\right)\,.
\end{split}
\label{Gn:bgk}
\end{equation}
Kazakov~\cite{K:85} obtained an expression for $G(1,1,1,1,n)$
via two ${}_3 F_2$ functions of argument $-1$ much earlier.
Recently, $G(n_1,n_2,n_3,n_4,n_5)$ for arbitrary indices
was calculated~\cite{BW:03} in terms of a double integral
or several double series.
These series can be systematically expanded in $\varepsilon$ up to any order,
and coefficients are expressed via multiple $\zeta$-values (Sect.~\ref{S:zeta}).

In order to calculate the 2-loop diagram of Fig.~\ref{F:Q2t}
with a 1-loop insertion in the middle light line,
we need $G(1,1,1,1,n+\varepsilon)$.
It is easy to shift $n_5$ by $\pm1$ using the relation~\cite{CT:81}
\begin{equation}
\left[ (d-2n_5-4)\5+ + 2(d-n_5-3) \right] G(1,1,1,1,n_5)
= 2\,\1+ (\3- - \2-\5+) G(1,1,1,1,n_5)\,,
\label{Gn:n5}
\end{equation}
which follows from integration-by-parts relations (Sect.~\ref{S:Q2})
(all terms in its right-hand side are trivial).
Therefore, it is sufficient to find it just for one $n$.
The simplest choice for which the algorithm of $\varepsilon$-expansion
(Sect.~\ref{S:hyper}) works for~(\ref{Gn:F}) is $n_5=2+\varepsilon$:
\begin{equation}
\begin{split}
&G(1,1,1,1,2+\varepsilon) =
\frac{2\Gamma(1+3\varepsilon)\Gamma(1-\varepsilon)}{1+2\varepsilon}\times{}\\
&\left[
\frac{\Gamma(1-2\varepsilon)\Gamma(1-\varepsilon)}%
{(2+\varepsilon)(1+\varepsilon)^2\Gamma(1-4\varepsilon)\Gamma(1+\varepsilon)}
{}_3 F_2 \left(
\begin{array}{c}
1,2-2\varepsilon,2+2\varepsilon\\3+\varepsilon,3+2\varepsilon
\end{array}
\right| \left.\vphantom{\frac{1}{1}}1\right)
+ \frac{\pi\cot3\pi\varepsilon}{2\varepsilon(1-2\varepsilon)}
\right]\,.
\end{split}
\label{Gn:F2}
\end{equation}
Expansion of this ${}_3 F_2$ in $\varepsilon$ is (Sect.~\ref{S:hyper})
\begin{gather}
{}_3 F_2 \left(
\begin{array}{c}
1,2-2\varepsilon,2+2\varepsilon\\3+\varepsilon,3+2\varepsilon
\end{array}
\right| \left.\vphantom{\frac{1}{1}}1\right) =
4(\zeta_2-1)
+ 6(-4\zeta_3+3\zeta_2-1)\varepsilon
\label{Gn:Fe}\\
{} + 2(41\zeta_4-54\zeta_3+22\zeta_2-9)\varepsilon^2
+3(-124\zeta_5+24\zeta_2\zeta_3+123\zeta_4-88\zeta_3+30\zeta_2-8)\varepsilon^3
+ \cdots
\nonumber
\end{gather}
Similarly, we obtain from~(\ref{Gn:bgk}) for $n_5=1+\varepsilon$
\begin{equation}
\begin{split}
&\frac{3\varepsilon^3(1-2\varepsilon)\Gamma(1+\varepsilon)\Gamma(1-4\varepsilon)}%
{\Gamma^2(1-\varepsilon)\Gamma(1-2\varepsilon)\Gamma(1+3\varepsilon)}
G(1,1,1,1,1+\varepsilon)\\
&{} = \frac{4}{3}\; {}_3 F_2 \left(
\begin{array}{c}1,-\varepsilon,3\varepsilon\\1+\varepsilon,1+3\varepsilon\end{array}
\right|\left.\vphantom{\frac{1}{1}}1\right)
- \frac{\Gamma(1+\varepsilon)\Gamma(1-4\varepsilon)}{\Gamma(1-\varepsilon)\Gamma(1-2\varepsilon)}
\pi\varepsilon \cot3\pi\varepsilon
- 1\\
&{} = - \frac{4}{3}\; {}_3 F_2 \left(
\begin{array}{c}1,-\varepsilon,-3\varepsilon\\1+\varepsilon,1-3\varepsilon\end{array}
\right|\left.\vphantom{\frac{1}{1}}1\right)
+ \frac{\Gamma(1+\varepsilon)\Gamma(1-4\varepsilon)}{\Gamma(1-\varepsilon)\Gamma(1-2\varepsilon)}
\pi\varepsilon \cot\pi\varepsilon
+ \frac{1}{3}\\
&{} = - \frac{1}{2}\; {}_3 F_2 \left(
\begin{array}{c}1,4\varepsilon,2\varepsilon\\1+\varepsilon,1+2\varepsilon\end{array}
\right|\left.\vphantom{\frac{1}{1}}1\right)
+ \frac{\Gamma(1+\varepsilon)\Gamma(1-4\varepsilon)}{\Gamma(1-\varepsilon)\Gamma(1-2\varepsilon)}
\pi\varepsilon \cot\pi\varepsilon
- \frac{1}{2}\\
&{} = \frac{1}{2}\; {}_3 F_2 \left(
\begin{array}{c}1,-\varepsilon,-2\varepsilon\\1-4\varepsilon,1-2\varepsilon\end{array}
\right|\left.\vphantom{\frac{1}{1}}1\right)
+ \frac{\Gamma(1+\varepsilon)\Gamma(1-4\varepsilon)}{\Gamma(1-\varepsilon)\Gamma(1-2\varepsilon)}
\pi\varepsilon \cot2\pi\varepsilon
- 1\,,
\end{split}
\label{Gn:bgk1}
\end{equation}
where
\begin{equation}
\begin{split}
&{}_3 F_2 \left(
\begin{array}{c}1,-\varepsilon,3\varepsilon\\1+\varepsilon,1+3\varepsilon\end{array}
\right|\left.\vphantom{\frac{1}{1}}1\right) =
1 - 3 \zeta_2 \varepsilon^2 + 18 \zeta_3 \varepsilon^3
- \frac{123}{2} \zeta_4 \varepsilon^4
+ 9 (31\zeta_5-6\zeta_2\zeta_3) \varepsilon^5
+ \cdots\\
&{}_3 F_2 \left(
\begin{array}{c}1,-\varepsilon,-3\varepsilon\\1+\varepsilon,1-3\varepsilon\end{array}
\right|\left.\vphantom{\frac{1}{1}}1\right) =
1 + 3 \zeta_2 \varepsilon^2 + \frac{69}{2} \zeta_4 \varepsilon^4
+ 27 (-3\zeta_5+2\zeta_2\zeta_3) \varepsilon^5 + \cdots\\
&{}_3 F_2 \left(
\begin{array}{c}1,4\varepsilon,2\varepsilon\\1+\varepsilon,1+2\varepsilon\end{array}
\right|\left.\vphantom{\frac{1}{1}}1\right) =
1 + 8 \zeta_2 \varepsilon^2 + 92 \zeta_4 \varepsilon^4
+ 72 (-3\zeta_5+2\zeta_2\zeta_3) \varepsilon^5 + \cdots\\
&\rlap{$\displaystyle{}_3 F_2 \left(
\begin{array}{c}1,-\varepsilon,-2\varepsilon\\1-4\varepsilon,1-2\varepsilon\end{array}
\right|\left.\vphantom{\frac{1}{1}}1\right) =
1 + 2 \zeta_2 \varepsilon^2 + 18 \zeta_3 \varepsilon^3 + 101 \zeta_4 \varepsilon^4
+ 18 (23\zeta_5+2\zeta_2\zeta_3) \varepsilon^5 + \cdots$}
\end{split}
\label{Gn:Fe1}
\end{equation}
Also, the last formula in~(\ref{Gn:bgk}) gives for $n_5=\varepsilon$
\begin{equation}
\begin{split}
G(1,1,1,1,\varepsilon) ={}&
\frac{2\Gamma(1-\varepsilon)\Gamma(1+3\varepsilon)}%
{3\varepsilon(1-2\varepsilon)(1-3\varepsilon)(1-4\varepsilon)
\Gamma(1+\varepsilon)\Gamma(1-4\varepsilon)}\\
&{}\times\Biggl[ \frac{1-\varepsilon}{1-2\varepsilon}
\Gamma(1-\varepsilon) \Gamma(1-2\varepsilon)
{}_3 F_2 \left(
\begin{array}{c}1,-\varepsilon,1-2\varepsilon\\2-4\varepsilon,2-2\varepsilon\end{array}
\right|\left.\vphantom{\frac{1}{1}}1\right)\\
&\hphantom{{}\times\Biggl[\Biggr.}{}
+ (1-4\varepsilon) \Gamma(1+\varepsilon) \Gamma(1-4\varepsilon) \pi \cot2\pi\varepsilon
- \Gamma(1-\varepsilon) \Gamma(1-2\varepsilon) \Biggr]\,,
\end{split}
\label{Gn:bgk0}
\end{equation}
where
\begin{gather}
{}_3 F_2 \left(
\begin{array}{c}1,-\varepsilon,1-2\varepsilon\\2-4\varepsilon,2-2\varepsilon\end{array}
\right|\left.\vphantom{\frac{1}{1}}1\right) =
1 + (\zeta_2-2) \varepsilon + (9\zeta_3-5\zeta_2-3) \varepsilon^2
\label{Gn:Fe0}\\
{} + \left(\frac{101}{2}\zeta_4-45\zeta_3+3\zeta_2-6\right) \varepsilon^3
+ \left(207\zeta_5+18\zeta_2\zeta_3-\frac{505}{2}\zeta_4+27\zeta_3+3\zeta_2-15\right)
\varepsilon^4 + \cdots
\nonumber
\end{gather}
It is not difficult to find several additional terms to~(\ref{Gn:Fe}),
(\ref{Gn:Fe1}), (\ref{Gn:Fe0}).
Of course, all 4 series~(\ref{Gn:bgk1}) for $G(1,1,1,1,1+\varepsilon)$ agree,
and those for $G(1,1,1,1,2+\varepsilon)$~(\ref{Gn:F2})
and $G(1,1,1,1,\varepsilon)$~(\ref{Gn:bgk0})
agree with the recurrence relation~(\ref{Gn:n5}).

\subsection{Non-planar basis integral}
\label{S:NP}

The non-planar basis diagram (the last one in Fig.~\ref{F:Q3b}) is finite;
its value at $\varepsilon=0$ can be obtained without calculations
by the method called gluing~\cite{CT:81}.

\begin{figure}[ht]
\begin{center}
\begin{picture}(70,82)
\put(21,69.5){\makebox(0,0){\includegraphics{q3t2.eps}}}
\put(43,69.5){\makebox(0,0)[l]{{$\displaystyle{}=\frac{20\zeta_5+\mathcal{O}(\varepsilon)}%
{(-p^2)^{2+3\varepsilon}}$}}}
\put(21,41){\makebox(0,0){\includegraphics{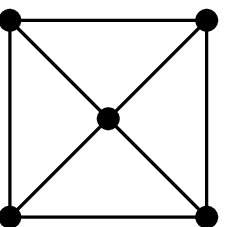}}}
\put(43,41){\makebox(0,0)[l]{{$\displaystyle{}=20\zeta_5\cdot\frac{1}{4\varepsilon}+\mathcal{O}(1)$}}}
\put(21,12.5){\makebox(0,0){\includegraphics{q3t3.eps}}}
\put(43,12.5){\makebox(0,0)[l]{{$\displaystyle{}=\frac{20\zeta_5+\mathcal{O}(\varepsilon)}%
{(-p^2)^{2+3\varepsilon}}$}}}
\end{picture}
\end{center}
\caption{Gluing}
\label{F:Glue}
\end{figure}

Let's consider the 4-loop vacuum diagram in the middle of Fig.~\ref{F:Glue}.
Let all lines have mass $m$, then there are no infrared divergences.
There are no divergent subdiagrams;
the diagram has an overall ultraviolet divergence,
and hence a $1/\varepsilon$ pole.

We can imagine that the middle vertex consists of 2 vertices
connected by a shrunk line (with power 0),
one vertex involves 2 lines on the left,
and the other one -- 2 lines on the right.
If we cut this shrunk line, we get the ladder diagram
(the upper one in Fig.~\ref{F:Glue}).
It is finite.
Therefore, the ultraviolet divergence of the vacuum diagram
comes from the last integration in $d^d p$ of the ladder diagram.
At $p\to\infty$, we may neglect $m$;
by dimensionality, the massless ladder diagram
behaves as $(A+\mathcal{O}(\varepsilon)/(-p^2)^{2+3\varepsilon}$.
Reducing it to the basis integrals and using results of Sect.~\ref{S:Gn},
we can obtain $A=20\zeta_5$
(first line of Fig.~\ref{F:Glue})%
\footnote{this $\mathcal{O}(1)$ term was found in~\cite{CT:81}
using Gegenbauer polynomials in $x$-space~\cite{CKT:80};
however, it is difficult to obtain further terms of $\varepsilon$ expansion
by this method.}.
It is not difficult to obtain several additional terms.
The ultraviolet divergence of the vacuum diagram comes from
\begin{equation*}
- \frac{i}{\pi^{d/2}} \left. \int \frac{d^d p}{(-p^2)^{2+3\varepsilon}} \right|_{\text{UV}}
= \frac{2}{\Gamma(2-\varepsilon)} \int_\lambda^\infty p_E^{-1-8\varepsilon} d p_E
= \frac{2}{\Gamma(2-\varepsilon)} \frac{\lambda^{-8\varepsilon}}{8\varepsilon}
= \frac{1}{4\varepsilon} + \mathcal{O}(1)\,,
\end{equation*}
where $\lambda$ is an infrared cutoff,
see the middle line in Fig.~\ref{F:Glue}.

On the other hand, we can imagine that one of the vertices in the middle of the vacuum diagram
involves the upper left line and the lower right one,
and the other vertex -- the lower left line and the upper right one.
Cutting the shrunk line, we get the non-planar diagram
(the lower one in Fig.~\ref{F:Glue}).
It is finite; at $p\to\infty$, where $m$ can be neglected,
it behaves as $(B+\mathcal{O}(\varepsilon)/(-p^2)^{2+3\varepsilon}$.
Integrating it in $d^d p$, we obtain the same vacuum diagram,
with the same $1/\varepsilon$ pole.
Therefore, $B=A$:
\emph{the non-planar basis diagram has the same value at $d=4$ as the ladder one}%
~\cite{CT:81}, namely, $20\zeta_5$ (Fig.~\ref{F:Glue}).
This method tells us nothing about further terms of expansion in $\varepsilon$.
The highly non-trivial problem of calculating the $\mathcal{O}(\varepsilon)$ term
in the non-planar basis diagram was solved in~\cite{K:84}.

\FloatBarrier
\section{HQET propagator diagrams}
\label{S:HQET}

\subsection{Crash course of HQET}
\label{S:Crash}

Heavy Quark Effective Theory
(HQET, see, e.g., \cite{Ne:94,Sh:99,MW:00,G:03})
is an effective field theory
constructed to approximate results of QCD for certain problems
with a single heavy quark.
Let the heavy quark stay approximately at rest (in some frame);
its momentum and energy are $|\vec{p}\,|\lesssim E$, $|p_0-m|\lesssim E$.
Momenta and energies of light quarks and gluons are
$|\vec{k}_i|\lesssim E$, $|k_{0i}|\lesssim E$.
Here $E\ll m$ is some fixed characteristic momentum,
and we consider the limit $m\to\infty$.
Scattering amplitudes and on-shell matrix elements of operators in QCD,
expanded up to some order in $E/m$, can be reproduced
from a simpler theory -- HQET.
The lowest-energy state of this theory (``vacuum'')
is a single heavy quark at rest, and has energy $m$.
Therefore, for any system containing this quark it is natural
to measure energy relative this zero level,
i.e., to consider its residual energy $\p_0=p_0-m$.
The free heavy quark has dispersion law
(dependence of energy on momentum, or mass shell)
$\p_0=\sqrt{m^2+\vec p\,^2}-m$.
Neglecting $1/m$ corrections, we may simplify it to
$\p_0(\vec{p}\,)=0$: the heavy-quark energy is zero,
and does not depend on its momentum.
The heavy quark at rest is described by a 2-component spinor field $Q$,
or a 4-component spinor having only upper components: $\gamma_0 Q=Q$.
The dispersion law $\p_0(\vec{p}\,)=0$ follows from
the Lagrangian $L=Q^+ i\partial_0 Q$.
Reintroducing the interaction with the gluon field
by the requirement of gauge invariance,
we obtain the leading-order (in $1/m$) HQET Lagrangian
\begin{equation}
L = Q^+ iD_0 Q + \cdots
\label{Crash:L}
\end{equation}
where all light-field parts (denoted by dots)
are exactly the same as in QCD.
The heavy quark interacts with $A_0$:
creates the coulomb chromoelectric field,
and reacts to external chromoelectric fields.

The Lagrangian~(\ref{Crash:L}) gives the heavy quark propagator
\begin{equation}
S(\p) = \frac{1}{\p_0+i0}\,,\quad
S(x) = -i \theta(x_0) \delta(\vec{x},)\,.
\label{Crash:S}
\end{equation}
In the momentum space it depends only on $\p_0$ but not on $\vec{p}$,
because we have neglected the kinetic energy.
Therefore, in the coordinate space the heavy quark does not move.
The vertex is $igv^\mu t^a$,
where $v^\mu=(1,0,0,0)$ is the heavy-quark 4-velocity.
Heavy-quark loops vanish, because it propagates only forward in time.

HQET is not Lorentz-invariant, because it has a preferred frame --
the heavy-quark rest frame.
However, it can be rewritten in covariant notations.
Momentum of any system containing the heavy quark is decomposed as
\begin{equation}
p = mv + \p\,,
\label{Crash:p}
\end{equation}
where the residual momentum is small: $|\p^\mu|\lesssim E$.
The heavy-quark field obeys $\rlap/v Q_v=Q_v$.
The Lagrangian is
\begin{equation}
L = \bar{Q}_v i v\cdot D Q_v + \cdots
\label{Crash:Lv}
\end{equation}
It gives the propagator
\begin{equation}
S(\p) = \frac{1+\rlap/v}{2} \frac{1}{\p\cdot v+i0}
\label{Crash:Sv}
\end{equation}
and the vertex $igv^\mu t^a$.

The QCD propagator at large $m$ becomes
\begin{equation}
S(p) = \frac{\rlap/p+m}{p^2-m^2}
= \frac{m(1+\rlap/v)+\rlap/\p}{2m\p\cdot v+\p\,^2}
= \frac{1+\rlap/v}{2}\frac{1}{\p\cdot v} + \mathcal{O}\left(\frac{\p}{m}\right)\,.
\label{HQET:QCDprop}
\end{equation}
A vertex $ig\gamma^\mu t^a$ sandwiched between two projectors
$\frac{1+\rlap/v}{2}$ may be replaced by $ig v^\mu t^a$.
Therefore, at the tree level all QCD diagrams become
the corresponding HQET diagrams, up to $1/m$ corrections.
In loop diagrams, momenta can be arbitrarily large,
and this correspondence breaks down.
Renormalization properties of HQET differ from those of QCD.
The ultraviolet behavior of an HQET diagram is determined
by the region of loop momenta much larger than the characteristic momentum scale
of the process $E$, but much less than the heavy quark mass $m$
(which tends to infinity from the very beginning).
It has nothing to do with the ultraviolet behavior
of the corresponding QCD diagram,
which is determined by the region of loop momenta much larger than $m$.

\subsection{1 loop}
\label{S:H1}

\begin{figure}[ht]
\begin{center}
\begin{picture}(64,27)
\put(32,13.5){\makebox(0,0){\includegraphics{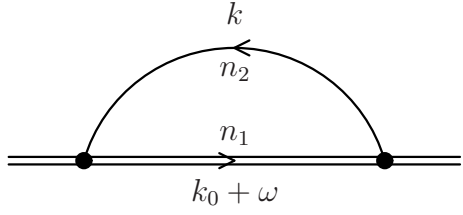}}}
\put(32,0){\makebox(0,0)[b]{{$k_0+\omega$}}}
\put(32,27){\makebox(0,0)[t]{{$k$}}}
\put(32,8){\makebox(0,0)[b]{{$n_1$}}}
\put(32,19){\makebox(0,0)[t]{{$n_2$}}}
\end{picture}
\end{center}
\caption{1-loop HQET propagator diagram}
\label{F:H1}
\end{figure}

The 1-loop HQET propagator diagram (Fig.~\ref{F:H1}) is
\begin{equation}
\begin{split}
&\int \frac{d^d k}{D_1^{n_1}D_2^{n_2}} =
i \pi^{d/2} (-2\omega)^{d-2n_2} I(n_1,n_2)\,,\\
&D_1 = \frac{k_0+\omega}{\omega}\,,\quad
D_2 = -k^2\,.
\end{split}
\label{H1:I}
\end{equation}
It vanishes if $n_1\le0$ or $n_2\le0$.

The Fourier transform of the 1-dimensional (HQET) propagator is
\begin{gather}
\int_{-\infty}^{+\infty} \frac{e^{-i\omega t}}{(-\omega-i0)^n}
\frac{d\omega}{2\pi} =
\frac{i^n}{\Gamma(n)} t^{n-1} \theta(t)\,,
\label{H1:p2x}\\
\int_0^\infty e^{i\omega t} t^n\, d t =
\frac{(-i)^{n+1} \Gamma(n+1)}{(-\omega-i0)^{n+1}}\,.
\label{H1:x2p}
\end{gather}
Our diagram in $x$-space is the product
of the heavy propagator~(\ref{H1:p2x}) with the power $n_1$
and the light propagator~(\ref{Q1:p2x}) with the power $n_2$.
The heavy quark stays at rest: $\vec{x}=0$.
Therefore, $-x^2=(it)^2$ in~(\ref{Q1:p2x})%
\footnote{Why not $(-it)^2$?
The Wick rotation to the Euclidean $x$-space is $t=-it_E$.
There are no imaginary parts at $t_E>0$.}.
Our diagram in the $x$-space is
\begin{equation*}
- 2^{-2n_2} \pi^{-d/2}
\frac{\Gamma(d/2-n_2)}{\Gamma(n_1) \Gamma(n_2)}
(i t)^{n_1+2n_2-d-1} \theta(t)\,.
\end{equation*}
Transforming it back to $p$-space~(\ref{H1:x2p}),
we arrive at
\begin{equation}
I(n_1,n_2) =
\frac{\Gamma(-d+n_1+2n_2) \Gamma(d/2-n_2)}{\Gamma(n_1) \Gamma(n_2)}\,.
\label{H1:I1}
\end{equation}

\subsection{2 loops}
\label{S:H2}

\begin{figure}[p]
\begin{center}
\begin{picture}(64,27)
\put(32,16.5){\makebox(0,0){\includegraphics{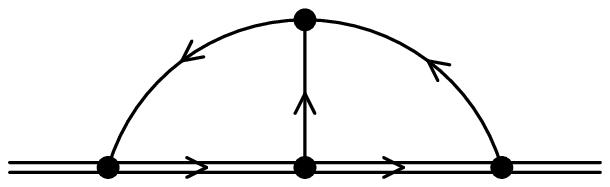}}}
\put(16,22){\makebox(0,0){{$k_1$}}}
\put(48,22){\makebox(0,0){{$k_2$}}}
\put(22,5){\makebox(0,0){{$k_{10}+\omega$}}}
\put(42,5){\makebox(0,0){{$k_{20}+\omega$}}}
\put(41,15){\makebox(0,0){{$k_1-k_2$}}}
\put(22,11.5){\makebox(0,0){{$n_1$}}}
\put(42,11.5){\makebox(0,0){{$n_2$}}}
\put(24,19){\makebox(0,0){{$n_3$}}}
\put(40,19){\makebox(0,0){{$n_4$}}}
\put(28,15){\makebox(0,0){{$n_5$}}}
\end{picture}
\end{center}
\caption{2-loop HQET propagator diagram $I$}
\label{F:h2ti}
\end{figure}

There are 2 generic topologies of 2-loop HQET propagator diagrams.
The first one is (Fig.~\ref{F:h2ti}):
\begin{gather}
\int \frac{d^d k_1\,d^d k_2}{D_1^{n_1} D_2^{n_2} D_3^{n_3} D_4^{n_4} D_5^{n_5}} =
- \pi^d (-2\omega)^{2(d-n_3-n_4-n_5)} I(n_1,n_2,n_3,n_4,n_5)\,,
\label{H2:I}\\
D_1=\frac{k_{10}+\omega}{\omega}\,,\quad
D_2=\frac{k_{20}+\omega}{\omega}\,,\quad
D_3=-k_1^2\,,\quad
D_4=-k_2^2\,,\quad
D_5=-(k_1-k_2)^2\,.
\nonumber
\end{gather}
This diagram is symmetric with respect to $(1\leftrightarrow2,3\leftrightarrow4)$.
It vanishes if indices of any two adjacent lines are non-positive.

\begin{figure}[p]
\begin{center}
\begin{picture}(52,19)
\put(26,9){\makebox(0,0){\includegraphics{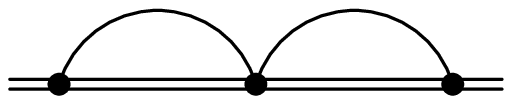}}}
\put(16,0){\makebox(0,0)[b]{{$n_1$}}}
\put(36,0){\makebox(0,0)[b]{{$n_2$}}}
\put(16,17){\makebox(0,0)[t]{{$n_3$}}}
\put(36,17){\makebox(0,0)[t]{{$n_4$}}}
\end{picture}
\end{center}
\caption{Trivial case $n_5=0$}
\label{F:H2z5}
\end{figure}

\begin{figure}[p]
\begin{center}
\begin{picture}(134,20)
\put(25,11.5){\makebox(0,0){\includegraphics{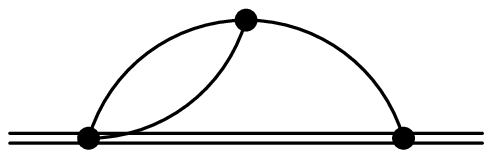}}}
\put(76,11.5){\makebox(0,0){\includegraphics{q1.eps}}}
\put(118,11.5){\makebox(0,0){\includegraphics{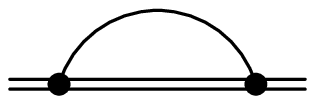}}}
\put(55,11.5){\makebox(0,0){{${}={}$}}}
\put(97,11.5){\makebox(0,0){{${}\times{}$}}}
\put(25,0){\makebox(0,0)[b]{{$n_2$}}}
\put(15,18){\makebox(0,0){{$n_3$}}}
\put(35,18){\makebox(0,0){{$n_4$}}}
\put(24,9){\makebox(0,0){{$n_5$}}}
\put(76,0){\makebox(0,0)[b]{{$n_5$}}}
\put(76,24){\makebox(0,0)[t]{{$\vphantom{d}n_3$}}}
\put(118,2){\makebox(0,0)[b]{{$n_2$}}}
\put(118,21){\makebox(0,0)[t]{{$n_4+n_3+n_5-d/2$}}}
\end{picture}
\end{center}
\caption{Trivial case $n_1=0$}
\label{F:H2z1}
\end{figure}

\begin{figure}[p]
\begin{center}
\begin{picture}(134,20)
\put(25,11.5){\makebox(0,0){\includegraphics{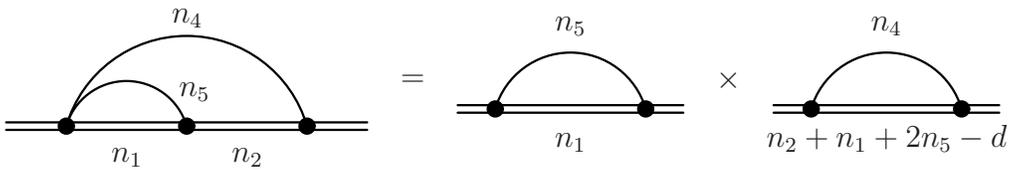}}}
\put(76,11.5){\makebox(0,0){\includegraphics{h1.eps}}}
\put(118,11.5){\makebox(0,0){\includegraphics{h1.eps}}}
\put(55,11.5){\makebox(0,0){{${}={}$}}}
\put(97,11.5){\makebox(0,0){{${}\times{}$}}}
\put(17,0){\makebox(0,0)[b]{{$n_1$}}}
\put(33,0){\makebox(0,0)[b]{{$n_2$}}}
\put(25,21){\makebox(0,0)[t]{{$n_4$}}}
\put(26,10){\makebox(0,0){{$n_5$}}}
\put(76,2){\makebox(0,0)[b]{{$n_1$}}}
\put(76,20){\makebox(0,0)[t]{{$n_5$}}}
\put(118,2){\makebox(0,0)[b]{{$n_2+n_1+2n_5-d$}}}
\put(118,20){\makebox(0,0)[t]{{$n_4$}}}
\end{picture}
\end{center}
\caption{Trivial case $n_3=0$}
\label{F:H2z3}
\end{figure}

If $n_5=0$, our diagram is just the product of two 1-loop diagrams
(Fig.~\ref{F:H2z5}):
\begin{equation}
I(n_1,n_2,n_3,n_4,0) = I(n_1,n_3) I(n_2,n_4)\,.
\label{H2:z5}
\end{equation}
If $n_1=0$, the inner loop gives $G(n_3,n_5)/(-k_2^2)^{n_3+n_5-d/2}$,
and (Fig.~\ref{F:H2z1})
\begin{equation}
I(0,n_2,n_3,n_4,n_5) = G(n_3,n_5) I(n_2,n_4+n_3+n_5-d/2)\,.
\label{H2:z1}
\end{equation}
If $n_3=0$, the inner loop gives $I(n_1,n_5)/(-k_0)^{n_1+2n_5-d}$,
and (Fig.~\ref{F:H2z3})
\begin{equation}
I(n_1,n_2,0,n_4,n_5) = I(n_1,n_5) I(n_2+n_1+2n_5-d,n_4)\,.
\label{H2:z3}
\end{equation}
The cases $n_2=0$, $n_4=0$ are symmetric.

When all $n_i>0$, we shall use integration by parts~\cite{BG:91}.
The differential operator $\partial/\partial k_2$
applied to the integrand of~(\ref{H2:I}) acts as
\begin{equation}
\frac{\partial}{\partial k_2} \to
-\frac{n_2}{D_2}\frac{v}{\omega} + \frac{n_4}{D_4}2k_2 + \frac{n_5}{D_5}2(k_2-k_1)
\label{H2:dk1}
\end{equation}
Applying $(\partial/\partial k_2)\cdot k_2$,
$(\partial/\partial k_2)\cdot(k_2-k_1)$ to the integrand of~(\ref{H2:I}),
we get vanishing integrals.
On the other hand, using
$k_2v/\omega=D_2-1$, $2(k_2-k_1)\cdot k_2=D_3-D_4-D_5$,
we see that these differential operators are equivalent to inserting
\begin{gather*}
d-n_2-n_5-2n_4 + \frac{n_2}{D_2} + \frac{n_5}{D_5}(D_3-D_4)\,,\\
d-n_2-n_4-2n_5 + \frac{n_2}{D_2}D_1 + \frac{n_4}{D_4}(D_3-D_5)
\end{gather*}
under the integral sign.
We obtain the recurrence relations
\begin{gather}
\left[ d-n_2-n_5-2n_4 + n_2\2+ + n_5\5+(\3--\4-) \right] I = 0\,,
\label{H2:Tri1}\\
\left[ d-n_2-n_4-2n_5 + n_2\2+\1- + n_4\4+(\3--\5-) \right] I = 0
\label{H2:Tri2}
\end{gather}
(two more relations are obtained by $(1\leftrightarrow3,2\leftrightarrow4)$).
Similarly, applying the differential operator $(\partial/\partial k_2)\cdot v$
 is equivalent to inserting
\begin{equation*}
- 2\frac{n_2}{D_2} + \frac{n_4}{D_4} 4\omega^2 (D_2-1)
+ \frac{n_5}{D_5} 4\omega^2 (D_2-D_1)\,,
\end{equation*}
and we obtain
\begin{equation}
\left[ -2n_2\2+ + n_4\4+(\2--1) + n_5\5+(\2--\1-) \right] I = 0\,.
\label{H2:Tri3}
\end{equation}
(there is also the symmetric relation, of course).

We can obtain a relation from homogeneity of the integral~(\ref{H2:I}) in $\omega$.
Applying the operator $\omega\cdot d/d\omega$ to $\omega^{-n_1-n_2}I$~(\ref{H2:I}),
we see that it is equivalent to the factor $2(d-n_3-n_4-n_5)-n_1-n_2$.
On the other hand, explicit differentiation of
$(-\omega D_1)^{-n_1}(-\omega D_2)^{-n_2}$ gives $-n_1/D_1-n_2/D_2$.
Therefore,
\begin{equation}
\left[ 2(d-n_3-n_4-n_5)-n_1-n_2 + n_1\1+ + n_2\2+ \right] I = 0\,.
\label{H2:hom}
\end{equation}
This is nothing but the sum of the $(\partial/\partial k_2)\cdot k_2$
relation~(\ref{H2:Tri1}) and its mirror-symmetric
$(\partial/\partial k_1)\cdot k_1$ relation.
A useful relation can be obtained by subtracting $\1-$ shifted~(\ref{H2:hom})
from~(\ref{H2:Tri2}):
\begin{equation}
\begin{split}
\bigl[& d-n_1-n_2-n_4-2n_5+1
- \bigl(2(d-n_3-n_4-n_5)-n_1-n_2+1\bigr)\1-\\
&{} + n_4\4+(\3--\5-) \bigr] I = 0\,.
\end{split}
\label{H2:David}
\end{equation}

Expressing $I(n_1,n_2,n_3,n_4,n_5)$ from~(\ref{H2:Tri2})
\begin{equation}
I(n_1,n_2,n_3,n_4,n_5) =
- \frac{n_2\2+\1- + n_4\4+(\3--\5-)}{d-n_2-n_4-2n_5} I\,,
\label{H2:Tri}
\end{equation}
we see that the sum $n_1+n_3+n_5$ reduces by 1
(Fig.~\ref{F:red}).
Therefore, applying~(\ref{Q2:Tri}) sufficiently many times,
we can reduce an arbitrary $I$ integral with integer indices
to a combination of integrals
with $n_5=0$ (Fig.~\ref{F:H2z5}, (\ref{H2:z5})),
$n_1=0$ (Fig.~\ref{F:H2z1}, (\ref{H2:z1})),
$n_3=0$ (Fig.~\ref{F:H2z3}, (\ref{H2:z3})).
Of course, if $\max(n_2,n_4)<\max(n_1,n_3)$, it may be more efficient
to use the relation mirror-symmetric to~(\ref{H2:Tri2}).
The relation~(\ref{H2:David}) also can be used instead of~(\ref{H2:Tri2}).

\begin{figure}[ht]
\begin{center}
\begin{picture}(72,42)
\put(36,21){\makebox(0,0){\includegraphics{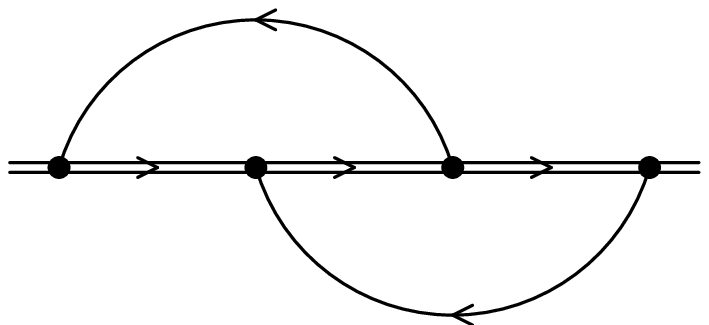}}}
\put(46,0){\makebox(0,0)[b]{{$k_2$}}}
\put(26,42){\makebox(0,0)[t]{{$k_1$}}}
\put(16,19.5){\makebox(0,0)[t]{{$k_{10}+\omega$}}}
\put(40,19.5){\makebox(0,0)[t]{{$k_{10}+k_{20}+\omega$}}}
\put(56,22.5){\makebox(0,0)[b]{{$k_{20}+\omega$}}}
\put(16,22.5){\makebox(0,0)[b]{{$n_1$}}}
\put(36,22.5){\makebox(0,0)[b]{{$n_3$}}}
\put(59,19.5){\makebox(0,0)[t]{{$\vphantom{k}n_2$}}}
\put(26,34){\makebox(0,0)[t]{{$n_4$}}}
\put(46,8){\makebox(0,0)[b]{{$n_5$}}}
\end{picture}
\end{center}
\caption{2-loop HQET propagator diagram $J$}
\label{F:h2tj}
\end{figure}

The second topology of 2-loop HQET propagator diagrams is (Fig.~\ref{F:h2tj}):
\begin{gather}
\int \frac{d^d k_1\,d^d k_2}{D_1^{n_1} D_2^{n_2} D_3^{n_3} D_4^{n_4} D_5^{n_5}} =
- \pi^d (-2\omega)^{2(d-n_4-n_5)} J(n_1,n_2,n_3,n_4,n_5)
\label{H2:J}\\
D_1=\frac{k_{10}+\omega}{\omega}\,,\quad
D_2=\frac{k_{20}+\omega}{\omega}\,,\quad
D_3=\frac{(k_1+k_2)_0+\omega}{\omega}\,,\quad
D_4=-k_1^2\,,\quad
D_5=-k_2^2\,.
\nonumber
\end{gather}
This diagram is symmetric with respect to $(1\leftrightarrow2,4\leftrightarrow5)$.
It vanishes if $n_4\le0$, or $n_5\le0$,
or two adjacent heavy indices are non-positive.
If $n_3=0$, our diagram is just the product of two 1-loop diagrams
(Fig.~\ref{F:H2z5}).
If $n_1=0$, it is also trivial (Fig.~\ref{F:H2z3}).

The denominators in~(\ref{H2:J}) are linearly dependent%
\footnote{HQET denominators are linear in momenta;
there are only 2 loop momenta, and 3 denominators cannot be independent.}:
$D_1+D_2-D_3=1$.
Therefore~\cite{BG:91}
\begin{equation}
J = (\1-+\2--\3-) J\,.
\label{H2:parfrac}
\end{equation}
Applying this relation sufficiently many times,
we can kill one of the lines 1, 2, 3,
and thus reduce any integral $J$ with integer indices to trivial cases.
In fact, we have not enough independent denominators
to express all scalar products in the numerator.
Therefore, we have to consider a more general integral than~(\ref{H2:J}),
containing powers of $k_1\cdot k_2$ in the numerator.
This wider class of integrals can also be reduced to the same
trivial cases~\cite{G:00}.

\begin{figure}[ht]
\begin{center}
\begin{picture}(84,17)
\put(16,8.5){\makebox(0,0){\includegraphics{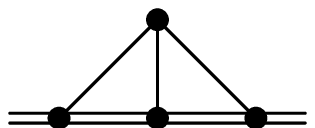}}}
\put(63,8.5){\makebox(0,0){\includegraphics{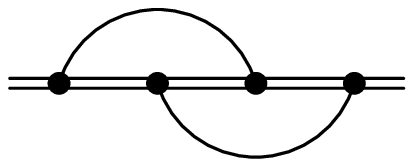}}}
\end{picture}
\end{center}
\caption{Topologies of 2-loop HQET propagator diagram}
\label{F:H2t}
\end{figure}

\begin{figure}[ht]
\begin{center}
\begin{picture}(104,10)
\put(11,5){\makebox(0,0){\includegraphics{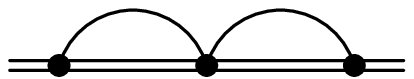}}}
\put(33,5){\makebox(0,0)[l]{{${}=I_1^2$}}}
\put(68,5){\makebox(0,0){\includegraphics{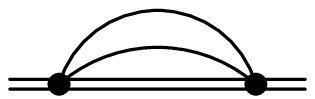}}}
\put(85,5){\makebox(0,0)[l]{{${}=I_2$}}}
\end{picture}
\end{center}
\caption{Basis diagrams (all indices equal 1)}
\label{F:H2b}
\end{figure}

\begin{figure}[ht]
\begin{center}
\begin{picture}(42,17)
\put(21,8.5){\makebox(0,0){\includegraphics{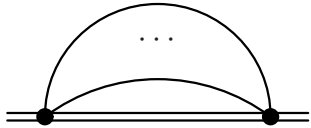}}}
\put(21,11){\makebox(0,0){{$\cdots$}}}
\end{picture}
\end{center}
\caption{$n$-loop HQET sunset diagrams}
\label{F:Hsun}
\end{figure}

Let's summarize.
There are 2 generic topologies of 2-loop HQET propagator diagrams
(Fig.~\ref{F:H2t}).
The integrals with all integer $n_i$
can be expressed as linear combinations of 2 basis integrals
(Fig.~\ref{F:H2b});
coefficients are rational functions of $d$.
Here $I_n$ are $n$-loop HQET sunset diagrams (Fig.~\ref{F:Hsun}):
\begin{equation}
I_n = \frac{\Gamma(1+2n\varepsilon)\Gamma^n(1-\varepsilon)}%
{(1-n(d-2))_{2n}}\,.
\label{H2:sun}
\end{equation}

\subsection{3 loops}
\label{S:H3}

\begin{figure}[ht]
\begin{center}
\begin{picture}(132,109)
\put(66,54.5){\makebox(0,0){\includegraphics{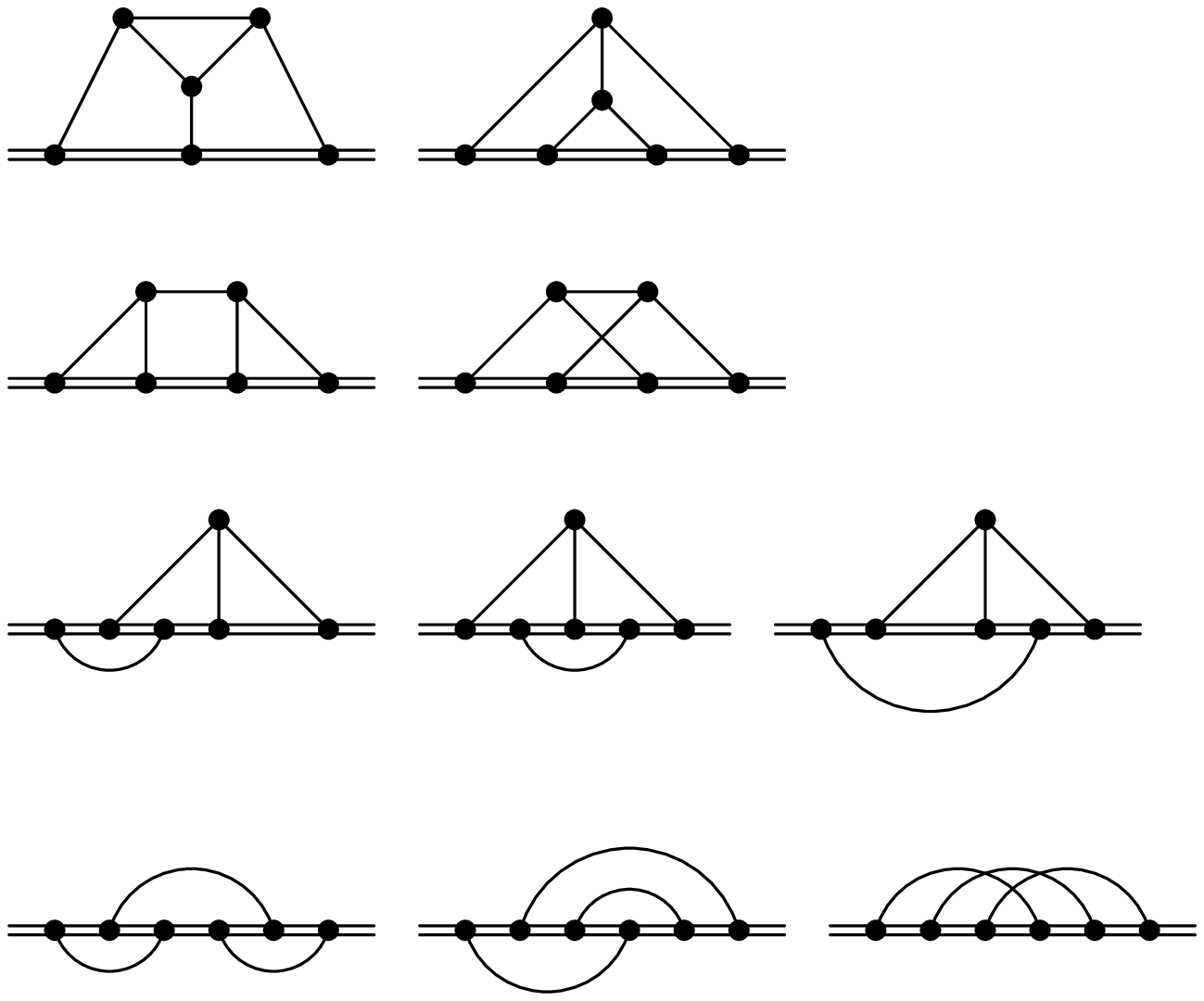}}}
\end{picture}
\end{center}
\caption{Topologies of 3-loop HQET propagator diagrams}
\label{F:H3t}
\end{figure}

\begin{figure}[ht]
\begin{center}
\begin{picture}(116,82)
\put(21,75){\makebox(0,0)[b]{\includegraphics{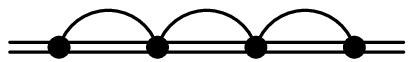}}}
\put(43,77){\makebox(0,0)[l]{{${}=I_1^3$}}}
\put(83,75){\makebox(0,0)[b]{\includegraphics{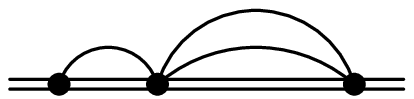}}}
\put(105,77){\makebox(0,0)[l]{{${}=I_1 I_2$}}}
\put(16,57){\makebox(0,0)[b]{\includegraphics{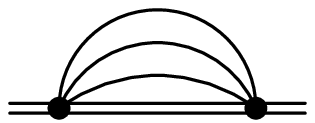}}}
\put(33,59){\makebox(0,0)[l]{{${}=I_3$}}}
\put(16,41){\makebox(0,0)[b]{\includegraphics{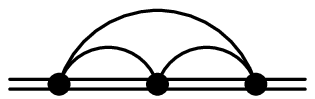}}}
\put(33,43){\makebox(0,0)[l]{{$\displaystyle{}\sim I_3\frac{I_1^2}{I_2}$}}}
\put(68,41){\makebox(0,0)[b]{\includegraphics{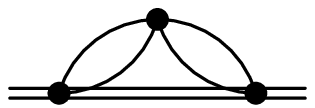}}}
\put(85,43){\makebox(0,0)[l]{{$\displaystyle{}\sim I_3\frac{G_1^2}{G_2}$}}}
\put(16,21){\makebox(0,0)[b]{\includegraphics{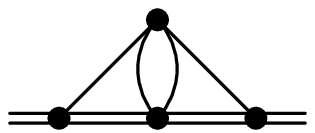}}}
\put(33,23){\makebox(0,0)[l]{{${}=G_1 I(1,1,1,1,\varepsilon)$}}}
\put(21,0){\makebox(0,0)[b]{\includegraphics{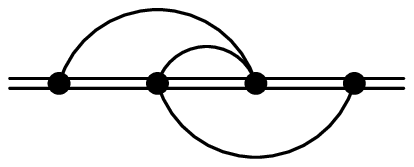}}}
\put(43,9){\makebox(0,0)[l]{{${}=I_1 J(1,1,-1+2\varepsilon,1,1)$}}}
\put(100,21){\makebox(0,0)[b]{\includegraphics{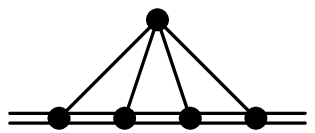}}}
\end{picture}
\end{center}
\caption{Basis diagrams (all indices equal 1, no numerators)}
\label{F:H3b}
\end{figure}

There are 10 generic topologies of 3-loop HQET propagator diagrams
(Fig.~\ref{F:H3t}).
Diagrams in the first 2 rows of the figure have one scalar product
which cannot be expressed via denominators;
those in the 3-rd row have 1 linear relation among heavy denominators,
and hence 2 independent scalar products in the numerator;
those in the last row have 2 relations among heavy denominators,
and hence 3 independent scalar products in the numerator.
Integration-by-parts recurrence relations for these diagrams
have been investigated in~\cite{G:00}.
They can be used to reduce all integrals of Fig.~\ref{F:H3t},
with arbitrary integer powers of denominators
and arbitrary numerators,
to linear combinations of 8 basis integrals (Fig.~\ref{F:H3b}).
This algorithm has been implemented in the \textsf{REDUCE}
package \textsf{Grinder}~\cite{G:00}%
\footnote{The hep-ph version of this paper contains some corrections
as compared to the journal version.}.
Recently, the heavy-quark propagator and the heavy-light current
anomalous dimension have been calculated~\cite{CG:03} at 3 loops,
using this package.

The first 5 basis integrals are trivial: they are expressed
via $I_n$~(\ref{H2:sun}) (Fig.~\ref{F:Hsun})
and $G_n$~(\ref{Q2:sun}) (Fig.~\ref{F:Qsun}),
and hence via $\Gamma$ functions.
The integrals in the 3-rd row differ from the one in the 2-nd row
by replacing a two-loop subdiagram.
In the first case, the 2-loop HQET subdiagram (Fig.~\ref{F:H2t}) is substituted:
$I_2/(-\omega)^{5-2d}\to I_1^2/(-\omega)^{6-2d}$.
In the second case, the 2-loop massless subdiagram (Fig.~\ref{F:Q2t}) is substituted:
$G_2/(-k^2)^{3-d}\to G_1^2/(-k^2)^{4-d}$.
Therefore, these diagrams are obtained from $I_3$ by multiplying by%
\footnote{\textsf{Grinder} uses $B_4=I_3 I_1^2/I_2$ and $B_5=I_3 G_1^2/G_2$
as elements of its basis, not the diagrams in Fig.~\ref{F:H3b}.}
\begin{gather*}
\frac{I_1^2 I(6-2d,1)}{I_2 I(5-2d,1)} =
\frac{3d-7}{2d-5} \frac{I_1^2}{I_2}\,,\\
\frac{G_1^2 I(1,4-d)}{G_2 I(1,3-d)} =
- 2 \frac{3d-7}{d-3} \frac{G_1^2}{G_2}\,.
\end{gather*}
Two basis integrals are proportional to $I(1,1,1,1,n)$ (Sect.~\ref{S:In})
and $J(1,1,n,1,1)$ (Sect.~\ref{S:Jn}), for non-integer $n$.
The last one is truly 3-loop and most difficult;
it will be discussed in Sect.~\ref{S:M3}.

\subsection{$J(1,1,n,1,1)$}
\label{S:Jn}

\begin{figure}[ht]
\begin{center}
\begin{picture}(42,17)
\put(21,8.5){\makebox(0,0){\includegraphics{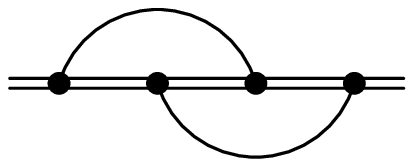}}}
\put(6,7){\makebox(0,0)[t]{$0$}}
\put(16,10){\makebox(0,0)[b]{{$t_1$}}}
\put(26,7){\makebox(0,0)[t]{{$t_2$}}}
\put(36,10){\makebox(0,0)[b]{{$\vphantom{{}_2}t$}}}
\put(21,10){\makebox(0,0)[b]{{$n$}}}
\end{picture}
\end{center}
\caption{$J(1,1,n,1,1)$ in $x$-space}
\label{F:Jt}
\end{figure}

This diagram (Fig.~\ref{F:Jt}) has been calculated in~\cite{G:00}.
In $x$-space, it is is
\begin{equation*}
\int_{0<t_1<t_2<t} dt_1\,dt_2\,
(i(t_2-t_1))^{n-1} (it_2)^{2-d} (i(t-t_1))^{2-d}\,,
\end{equation*}
or, going to Euclidean space ($t_i=-it_{Ei}$),
\begin{equation*}
\int_{0<t_1<t_2<t} dt_1\,dt_2\,
(t_2-t_1)^{n-1} t_2^{2-d} (t-t_1)^{2-d} =
J t^{n-2d+5}\,,
\end{equation*}
where the power of $t$ is obvious from counting dimensions.
Collecting factors from Fourier transforms, we have
\begin{equation}
J(1,1,n,1,1) =
\frac{\Gamma(n-2d+6)\Gamma^2(d/2-1)}{\Gamma(n)} J\,,
\label{Jn:Jn}
\end{equation}
where the dimensionless integral $J$ is
\begin{equation*}
J = \int_{0<t_1<t_2<1} dt_1\,dt_2\,
(t_2-t_1)^{n-1} t_2^{2-d} (1-t_1)^{2-d}\,.
\end{equation*}
The substitution $t_1=x t_2$ gives
\begin{equation*}
J = \int_0^1 dt\,t^{n-d+2}
\int_0^1 dx\,(1-x)^{n-1} (1-xt)^{2-d}\,.
\end{equation*}

Integrations in $t$ and $x$ would factor, if not the factor $(1-xt)^{2-d}$.
Let's expand this troublesome factor according to the Newton binomial:
\begin{equation*}
(1-xt)^{2-d} =
\sum_{k=0}^\infty \frac{(d-2)_k}{k!} (xt)^k\,.
\end{equation*}
Here the Pochhammer symbol is
\begin{equation}
(x)_k = \prod_{i=0}^{k-1} (x+i) = \frac{\Gamma(x+k)}{\Gamma(x)}\,.
\label{Jn:Poch}
\end{equation}
Then the integrations become trivial:
\begin{equation}
J = \Gamma(n) \sum_{k=0}^\infty
\frac{(d-2)_k}{(n-d+k+3)\Gamma(n+k+1)}
= \frac{1}{n(n-d+3)} \sum_{k=0}^\infty
\frac{(d-2)_k (n-d+3)_k}{(n+1)_k (n-d+4)_k}\,.
\label{Jn:Js}
\end{equation}
Recalling the definition of the hypergeometric function
\begin{equation}
{}_3 F_2\left(
\begin{array}{c}a_1,a_2,a_3\\b_1,b_2\end{array}
\right|\left.\vphantom{\frac{1}{1}}x\right) =
\sum_{k=0}^\infty
\frac{(a_1)_k (a_2)_k (a_3)_k}{(b_1)_k (b_2)_k}
\frac{x^k}{k!}
\label{Jn:Hyper}
\end{equation}
and taking into account $(1)_k=k!$,
we can rewrite the result as
\begin{equation*}
J = \frac{1}{n(n-d+3)}\,_3 F_2 \left(
\begin{array}{c}1,d-2,n-d+3\\n+1,n-d+4\end{array}
\right|\left.\vphantom{\frac{1}{1}}1\right)\,,
\end{equation*}
and hence~(\ref{Jn:Jn})
\begin{equation}
J(1,1,n,1,1) =
\frac{\Gamma(n-2d+6)\Gamma^2(d/2-1)}{(n-d+3)\Gamma(n+1)}
{}_3 F_2 \left(
\begin{array}{c}1,d-2,n-d+3\\n+1,n-d+4\end{array}
\right|\left.\vphantom{\frac{1}{1}}1\right)\,.
\label{Jn:JF}
\end{equation}
This is the easiest diagram calculation in the world
having ${}_3 F_2$ as its result:
just a double integral.

It is not more difficult to obtain the general result%
\footnote{I've presented only the result for $n_1=n_2=1$ in~\cite{G:00},
I've no idea why: the general case is not more difficult.}
\begin{equation}
\begin{split}
&J(n_1,n_2,n_3,n_4,n_5) = {}\\
&\frac{\Gamma(n_1+n_2+n_3+2(n_4+n_5-d))\Gamma(n_1+n_3+2n_4-d)
\Gamma(d/2-n_4)\Gamma(d/2-n_5)}%
{\Gamma(n_1+n_2+n_3+2n_4-d)\Gamma(n_4)\Gamma(n_5)\Gamma(n_1+n_3)}\\
&{}\times{}_3F_2 \left(
\begin{array}{c}
n_1,d-2n_5,n_1+n_3+2n_4-d\\
n_1+n_3,n_1+n_2+n_3+2n_4-d
\end{array}
\right|\left.\vphantom{\frac{1}{1}}1\right)\,.
\end{split}
\label{Jn:J}
\end{equation}

In order to calculate the 2-loop diagram of Fig.~\ref{F:h2tj}
with a 1-loop insertion in the middle heavy line,
we need $J(1,1,n+2\varepsilon,1,1)$.
It is easy to shift $n_3$ by $\pm1$ using~(\ref{H2:parfrac}).
Therefore, it is sufficient to find it just for one $n$.
The simplest choice for which the algorithm of $\varepsilon$-expansion
(Sect.~\ref{S:hyper}) works is $n=2$.
We obtain~\cite{CG:03} from~(\ref{Jn:JF})
\begin{equation}
\begin{split}
&J(1,1,2+2\varepsilon,1,1) =
\frac{1}{3(d-4)(d-5)(d-6)(2d-9)}\\
&{}\times
\frac{\Gamma(1+6\varepsilon)\Gamma^2(1-\varepsilon)}{\Gamma(1+2\varepsilon)}
\,_3 F_2 \left(
\begin{array}{c}
1,2-2\varepsilon,1+4\varepsilon\\
3+2\varepsilon,2+4\varepsilon
\end{array}
\right|\left.\vphantom{\frac{1}{1}}1\right)\,.
\end{split}
\label{Jn:J2}
\end{equation}
Expansion of this ${}_3 F_2$ in $\varepsilon$ is (Sect.~\ref{S:hyper})
\begin{equation}
\begin{split}
&{}_3 F_2 \left(
\begin{array}{c}
1,2-2\varepsilon,1+4\varepsilon\\
3+2\varepsilon,2+4\varepsilon
\end{array}
\right|\left.\vphantom{\frac{1}{1}}1\right) =
2 + 6(-2\zeta_2+3) \varepsilon\\
&{} + 12(10\zeta_3-11\zeta_2+6) \varepsilon^2
+ 24(-28\zeta_4+55\zeta_3-27\zeta_2+9) \varepsilon^3\\
&{} + 48(94\zeta_5-16\zeta_2\zeta_3-154\zeta_4+135\zeta_3-45\zeta_2+12) \varepsilon^4
+ \cdots
\end{split}
\label{Jn:F}
\end{equation}
It is not difficult to find several additional terms.

\subsection{$I(1,1,1,1,n)$}
\label{S:In}

\begin{figure}[ht]
\begin{center}
\begin{picture}(32,12)
\put(16,6){\makebox(0,0){\includegraphics{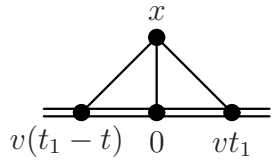}}}
\put(16,-5){\makebox(0,0)[b]{{$\vphantom{{}_1}0$}}}
\put(26,-5){\makebox(0,0)[b]{{$vt_1$}}}
\put(4,-5){\makebox(0,0)[b]{{$v(t_1-t)$}}}
\put(16,15){\makebox(0,0)[t]{{$x$}}}
\end{picture}
\end{center}
\caption{$I(1,1,1,1,n)$ in $x$-space}
\label{F:It}
\end{figure}

This diagram (Fig.~\ref{F:It}) has been calculated in~\cite{BB:94},
using Gegenbauer polynomials in $x$-space~\cite{CKT:80}.
This is, probably, the simplest example of applying this useful technique,
therefore, we shall consider some details of this calculation.
In Euclidean $x$-space, this diagram is
\begin{equation*}
\int_0^t dt_1 \int d^d x\,
\frac{1}{(x^2)^{d/2-n}
\left[(x-vt_1)^2\right]^\lambda
\left[(x+v(t-t_1))^2\right]^\lambda}
= \frac{2\pi^{d/2}}{\Gamma(d/2)} I t^{2(n-d)+5}\,,
\end{equation*}
where $\lambda=d/2-1$,
and the power of $t$ is obvious from counting dimensions.
The middle vertex has been chosen as the origin,
because the propagator with the non-standard power begins
at this vertex (it is much easier to handle propagators
beginning at the origin).
Collecting factors from Fourier transforms, we have
\begin{equation}
I(1,1,1,1,n) = \frac{2}{\pi}
\frac{\Gamma(2(n-d+3))\Gamma(d/2-n)\Gamma^2(d/2-1)}{\Gamma(d/2)\Gamma(n)} I\,.
\label{In:In}
\end{equation}
The dimensionless integral $I$ is
\begin{equation*}
I = \int_0^1 dt \int_0^\infty dx\,d\hat{x}\,
\frac{x^{2n-1}}{\left[(x-vt)^2\right]^\lambda
\left[(x+v(1-t))^2\right]^\lambda}\,,
\end{equation*}
where
\begin{equation*}
d^d x = \frac{2\pi^{d/2}}{\Gamma(d/2)} x^{d-1} dx\,d\hat{x}\,,
\end{equation*}
and the angular integration measure $d\hat{x}$ is normalized:
\begin{equation*}
\int d\hat{x} = 1\,.
\end{equation*}

This problem is more difficult than the one in Sect.~\ref{S:Jn}:
now we have $d$-dimensional integration over the coordinates of the light vertex.
It can be simplified by expanding 2 massless propagators
(whose ends are not at the origin) in series in Gegenbauer polynomials~\cite{CKT:80}
\begin{equation}
\frac{1}{\left[(x-y)^2\right]^\lambda} =
\frac{1}{\max^{2\lambda}(x,y)}
\sum_{k=0}^\infty T^k(x,y) C^\lambda_k(\hat{x}\cdot\hat{y})\,,
\label{In:Gegen}
\end{equation}
where
\begin{equation*}
T(x,y) = \min \left( \frac{x}{y}\,,\,\frac{y}{x} \right)\,.
\end{equation*}
Then the angular integration can be easily done using the orthogonality relation
\begin{equation}
\int d\hat{x}\,C^\lambda_{k_1}(\hat{a}\cdot\hat{x})
C^\lambda_{k_2}(\hat{b}\cdot\hat{x}) =
\delta_{k_1 k_2} \frac{\lambda}{\lambda+k_1}
C^\lambda_{k_1}(\hat{a}\cdot\hat{b})\,.
\label{In:Orth}
\end{equation}
In our particular case, the unit vectors $\hat{a}$ and $\hat{b}$
are $v$ and $-v$, and the result involves
\begin{equation*}
C^\lambda_k(-1) = (-1)^k C^\lambda_k(1) =
(-1)^k \frac{\Gamma(2\lambda+k)}{k!\,\Gamma(2\lambda)}\,.
\end{equation*}

Now we have a single sum:
\begin{gather*}
I = \frac{d-2}{\Gamma(d-2)} \int_0^1 dt
\sum_{k=0}^\infty \frac{(-1)^k}{k!} \frac{\Gamma(d+k-2)}{d+2k-2} I_k(t)\,,\\
I_k(t) = \int_0^\infty dx
\frac{x^{2n-1}\left[T(x,t)T(x,1-t)\right]^k}{\left[\max(x,t)\max(x,1-t)\right]^{d-2}}\,.
\end{gather*}
The contributions of the regions $t<\frac{1}{2}$ and $t>\frac{1}{2}$
are equal, and we may consider the first of them and double it.
The radial integral (in $x$) has to be calculated in several intervals
separately, because of $\max$ and $T$ in~(\ref{In:Gegen}):
\begin{equation*}
I_k(t) =
\int_0^t dx
\frac{x^{2n-1}\left[\frac{x}{t}\frac{x}{1-t}\right]^k}{\left[t(1-t)\right]^{d-2}}
+ \int_t^{1-t} dx
\frac{x^{2n-1}\left[\frac{t}{x}\frac{x}{1-t}\right]^k}{\left[x(1-t)\right]^{d-2}}
+ \int_{1-t}^\infty dx
\frac{x^{2n-1}\left[\frac{t}{x}\frac{1-t}{x}\right]^k}{\left[x^2\right]^{d-2}}\,.
\end{equation*}
The result is
\begin{align*}
I =&{} \frac{d-2}{(d-2n-2)\Gamma(d-2)}
\int_0^{1/2} dt \sum_{k=0}^\infty \frac{(-1)^k}{k!} \Gamma(d+k-2)\\
&{}\times\left[\frac{t^{-d+2n+k+2}(1-t)^{-d-k+2}}{n+k}
- \frac{t^k(1-t)^{-2d+2n-k+4}}{d-n+k-2}\right]\,.
\end{align*}

Unfortunately, the integral in $t$ cannot be immediately evaluated.
The authors of~\cite{BB:94} had to do some more juggling to obtain
\begin{gather}
I(1,1,1,1,n) = 2
\Gamma\left(\frac{d}{2}-1\right) \Gamma\left(\frac{d}{2}-n-1\right) \times{}
\label{In:F1}\\
\left[\frac{\Gamma(2n-2d+6)}{(2n-d+3)\Gamma(n+1)}
\,_3 F_2 \left(
\begin{array}{c}1,d-2,2n-d+3\\n+1,2n-d+4\end{array}
\right|\left.\vphantom{\frac{1}{1}}1\right)
- \frac{\Gamma(d-n-2)\Gamma^2(n-d+3)}{\Gamma(d-2)}
\right]\,.
\nonumber
\end{gather}
This result can also be rewritten~\cite{BB:94} as
\begin{equation}
\begin{split}
I(1,1,1,1,n) =
\frac{\Gamma\left(\frac{d}{2}-1\right)\Gamma\left(\frac{d}{2}-n-1\right)}%
{\Gamma(d-2)}
\Biggl[ 2
\frac{\Gamma(2n-d+3)\Gamma(2n-2d+6)}{(n-d+3)\Gamma(3n-2d+6)}\times{}\\
{}_3 F_2 \left(
\begin{array}{c}n-d+3,n-d+3,2n-2d+6\\n-d+4,3n-2d+6\end{array}
\right|\left.\vphantom{\frac{1}{1}}1\right)
- \Gamma(d-n-2) \Gamma^2(n-d+3) \Biggr]\,,
\end{split}
\label{In:F2}
\end{equation}
this series has larger region of convergence.

In order to calculate the 2-loop diagram of Fig.~\ref{F:h2ti}
with a 1-loop insertion in the middle light line,
we need $I(1,1,1,1,n+\varepsilon)$.
It is easy to shift $n_5$ by $\pm1$ using the relation~\cite{G:00}
\begin{equation}
\begin{split}
\left[(d-2n_5-4)\5+ - 2(d-n_5-3)\right] I(1,1,1,1,n_5) ={}\\
\left[(2d-2n_5-7)\1-\5+ - \3-\4+\5- + \1-\3+ \right] I(1,1,1,1,n_5)\,,
\end{split}
\label{In:n5}
\end{equation}
which follows from integration-by-parts relations (Sect.~\ref{S:H2})
(all terms in its right-hand side are trivial).
Therefore, it is sufficient to find it just for one $n$.
The simplest choice for which the algorithm of $\varepsilon$-expansion
(Sect.~\ref{S:hyper}) works is to use $n=1$ in~(\ref{In:F2}).
We obtain~\cite{CG:03}
\begin{equation}
\begin{split}
&I(1,1,1,1,1+\varepsilon) =
\frac{4\Gamma(1-\varepsilon)}{9(d-3)(d-4)^2}\times{}\\
&\Biggl[
\frac{\Gamma(1+4\varepsilon)\Gamma(1+6\varepsilon)}{\Gamma(1+7\varepsilon)}
\,_3 F_2 \left(
\begin{array}{c}
3\varepsilon,3\varepsilon,6\varepsilon\\
1+3\varepsilon,1+7\varepsilon
\end{array}
\right| \left. \vphantom{\frac{1}{1}} 1 \right)
- \Gamma^2(1+3\varepsilon) \Gamma(1-3\varepsilon) \Biggr]\,.
\end{split}
\label{In:I1}
\end{equation}
This ${}_3 F_2$ function has 3 upper indices $\sim\varepsilon$;
therefore, its expansion (apart from the leading 1)
starts from $\mathcal{O}(\varepsilon^3)$:
\begin{equation}
{}_3 F_2 \left(
\begin{array}{c}
3\varepsilon,3\varepsilon,6\varepsilon\\
1+3\varepsilon,1+7\varepsilon
\end{array}
\right| \left. \vphantom{\frac{1}{1}} 1 \right) =
1 + 54 \zeta_3 \varepsilon^3 - 513 \zeta_4 \varepsilon^4
+ 54 (25\zeta_5+28\zeta_2\zeta_3) \varepsilon^5 + \cdots
\label{In:F}
\end{equation}
It is not difficult to find several additional terms.

\FloatBarrier
\section{Massive on-shell propagator diagrams}
\label{S:M}

In this Section,
we shall consider self-energy diagrams of a massive particle with mass $m$,
when the external momentum is on the mass shell: $p^2=m^2$, or $p=mv$.
Why this special case is interesting?
First of all, in order to calculate any $S$-matrix elements,
one has to consider mass~\cite{GBGS:90,MR:00b}
and wave-function~\cite{BGS:91,MR:00c} renormalization
in the on-shell scheme.
Calculations of form factors and their derivatives at the point $q=0$
belong to this class, e.g., the anomalous magnetic moment~\cite{LR:96}
and electron charge radius~\cite{MR:00a} in QED.
Finally, such calculations are used for QCD/HQET matching~\cite{BG:95,CG:97}.

\subsection{1 loop}
\label{S:M1}

\begin{figure}[ht]
\begin{center}
\begin{picture}(64,27)
\put(32,13.5){\makebox(0,0){\includegraphics{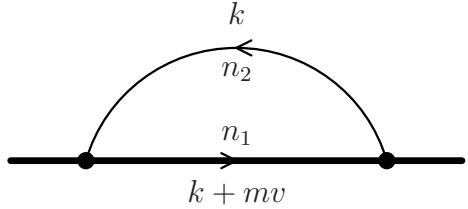}}}
\put(32,0){\makebox(0,0)[b]{{$k+mv$}}}
\put(32,27){\makebox(0,0)[t]{{$k$}}}
\put(32,8){\makebox(0,0)[b]{{$n_1$}}}
\put(32,19){\makebox(0,0)[t]{{$n_2$}}}
\end{picture}
\end{center}
\caption{1-loop on-shell propagator diagram}
\label{F:M1}
\end{figure}

The 1-loop on-shell propagator diagram (Fig.~\ref{F:M1}) is
\begin{equation}
\begin{split}
&\int \frac{d^d k}{D_1^{n_1}D_2^{n_2}} =
i \pi^{d/2} m^{d-2(n_1+n_2)} M(n_1,n_2)\,,\\
&D_1 = m^2-(k+mv)^2\,,\quad
D_2 = -k^2\,.
\end{split}
\label{M1:M}
\end{equation}
It vanishes if $n_1\le0$.

\begin{figure}[ht]
\begin{center}
\begin{picture}(74,19)
\put(16,9){\makebox(0,0){\includegraphics{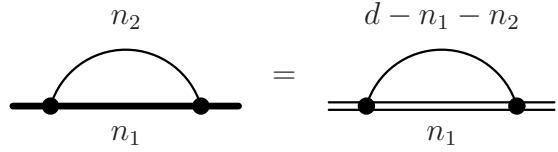}}}
\put(58,9){\makebox(0,0){\includegraphics{h1.eps}}}
\put(37,9){\makebox(0,0){{${}={}$}}}
\put(16,0){\makebox(0,0)[b]{{$n_1$}}}
\put(16,19){\makebox(0,0)[t]{{$\vphantom{d}n_2$}}}
\put(58,0){\makebox(0,0)[b]{{$n_1$}}}
\put(58,19){\makebox(0,0)[t]{{$d-n_1-n_2$}}}
\end{picture}
\end{center}
\caption{Inversion relation}
\label{F:M1i}
\end{figure}

The definition~(\ref{M1:M}) of $M(n_1,n_2)$ expressed via
the dimensionless Euclidean momentum $K=k_E/m$ becomes
\begin{equation*}
\int \frac{d^d K}{(K^2-2iK_0)^{n_1}(K^2)^{n_2}} =
\pi^{d/2} M(n_1,n_2)\,.
\end{equation*}
The definition~(\ref{H1:I}) of the 1-loop HQET integral $I(n_1,n_2)$
expressed via the dimensionless Euclidean momentum $K=k_E/(-2\omega)$ is
\begin{equation*}
\int \frac{d^d K}{(1-2iK_0)^{n_1}(K^2)^{n_2}} =
\pi^{d/2} I(n_1,n_2)\,.
\end{equation*}
Inversion~(\ref{Q1:Inv}) transforms the on-shell denominator
to the HQET one:
\begin{equation}
K^2 - 2 i K_0 = \frac{1 - 2 i K_0'}{K^{\prime2}}
\label{M1:Inv}
\end{equation}
(and vice versa).
Therefore, the problem of calculating $M(n_1,n_2)$
reduces to the previously solved one for $I(n_1,n_2)$ (Sect.~\ref{S:H1}):
\begin{equation}
M(n_1,n_2) = I(n_1,d-n_1-n_2) =
\frac{\Gamma(d-n_1-2n_2)\Gamma(-d/2+n_1+n_2)}{\Gamma(n_1)\Gamma(d-n_1-n_2)}
\label{M1:M1}
\end{equation}
(Fig.~\ref{F:M1i}).

\subsection{2 loops}
\label{S:M2}

\begin{figure}[p]
\begin{center}
\begin{picture}(64,27)
\put(32,16.5){\makebox(0,0){\includegraphics{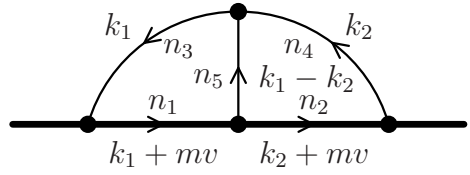}}}
\put(16,22){\makebox(0,0){{$k_1$}}}
\put(48,22){\makebox(0,0){{$k_2$}}}
\put(22,5){\makebox(0,0){{$k_1+mv$}}}
\put(42,5){\makebox(0,0){{$k_2+mv$}}}
\put(41,15){\makebox(0,0){{$k_1-k_2$}}}
\put(22,11.5){\makebox(0,0){{$n_1$}}}
\put(42,11.5){\makebox(0,0){{$n_2$}}}
\put(24,19){\makebox(0,0){{$n_3$}}}
\put(40,19){\makebox(0,0){{$n_4$}}}
\put(28,15){\makebox(0,0){{$n_5$}}}
\end{picture}
\end{center}
\caption{2-loop on-shell propagator diagram $M$}
\label{F:m2tm}
\end{figure}

\begin{figure}[p]
\begin{center}
\begin{picture}(72,42)
\put(36,21){\makebox(0,0){\includegraphics{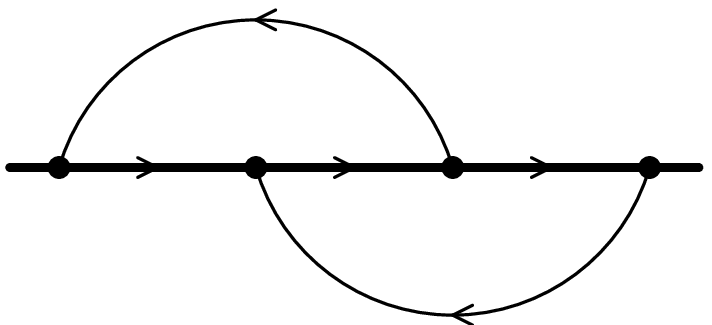}}}
\put(46,0){\makebox(0,0)[b]{{$k_2$}}}
\put(26,42){\makebox(0,0)[t]{{$k_1$}}}
\put(16,19.5){\makebox(0,0)[t]{{$k_1+mv$}}}
\put(40,19.5){\makebox(0,0)[t]{{$k_1+k_2+mv$}}}
\put(56,22.5){\makebox(0,0)[b]{{$k_2+mv$}}}
\put(16,22.5){\makebox(0,0)[b]{{$n_1$}}}
\put(36,22.5){\makebox(0,0)[b]{{$n_3$}}}
\put(59,19.5){\makebox(0,0)[t]{{$\vphantom{k}n_2$}}}
\put(26,34){\makebox(0,0)[t]{{$n_4$}}}
\put(46,8){\makebox(0,0)[b]{{$n_5$}}}
\end{picture}
\end{center}
\caption{2-loop on-shell propagator diagram $N$}
\label{F:m2tn}
\end{figure}

There are 2 generic topologies of 2-loop on-shell propagator diagrams.
The first one is (Fig.~\ref{F:m2tm}):
\begin{gather}
\int \frac{d^d k_1\,d^d k_2}{D_1^{n_1} D_2^{n_2} D_3^{n_3} D_4^{n_4} D_5^{n_5}} =
- \pi^d m^{2(d-\sum n_i)} M(n_1,n_2,n_3,n_4,n_5)\,,
\label{M2:M}\\
D_1=m^2-(k_1+mv)^2\,,\quad
D_2=m^2-(k_2+mv)^2\,,
\nonumber\\
D_3=-k_1^2\,,\quad
D_4=-k_2^2\,,\quad
D_5=-(k_1-k_2)^2\,.
\nonumber
\end{gather}
The second topology is (Fig.~\ref{F:m2tn}):
\begin{gather}
\int \frac{d^d k_1\,d^d k_2}{D_1^{n_1} D_2^{n_2} D_3^{n_3} D_4^{n_4} D_5^{n_5}} =
- \pi^d m^{2(d-\sum n_i)} N(n_1,n_2,n_3,n_4,n_5)\,,
\label{M2:N}\\
D_1=m^2-(k_1+mv)^2\,,\quad
D_2=m^2-(k_2+mv)^2\,,\quad
D_3=m^2-(k_1+k_2+mv)^2\,,
\nonumber\\
D_4=-k_1^2\,,\quad
D_5=-k_2^2\,.
\nonumber
\end{gather}

\begin{figure}[p]
\begin{center}
\begin{picture}(112,26)
\put(21,13){\makebox(0,0){\includegraphics{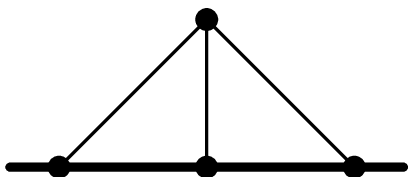}}}
\put(82,13){\makebox(0,0){\includegraphics{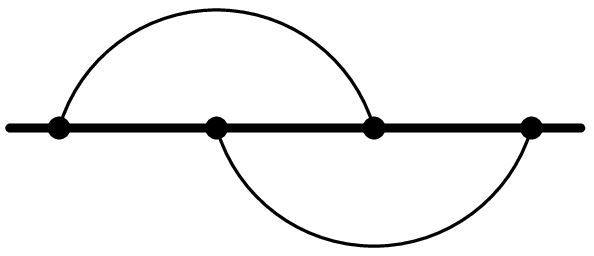}}}
\end{picture}
\end{center}
\caption{Topologies of 2-loop on-shell propagator diagrams}
\label{F:M2t}
\end{figure}

\begin{figure}[p]
\begin{center}
\begin{picture}(87,17)
\put(43.5,8.5){\makebox(0,0){\includegraphics{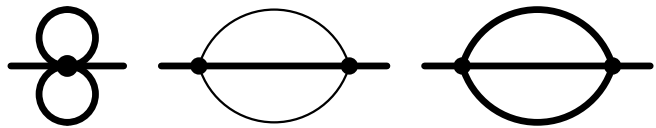}}}
\end{picture}
\end{center}
\caption{Basis diagrams (all indices equal 1)}
\label{F:M2b}
\end{figure}

The integrals of both generic topologies (Fig.~\ref{F:M2t})
with all integer $n_i$ can be reduced,
by using integration-by-parts recurrence relations,
to linear combinations of 3 basis integrals (Fig.~\ref{F:M2b});
coefficients are rational functions of $d$.
Only the topology $N$ involve the last, and the most difficult,
basis integral in Fig.~\ref{F:M2b}.
The reduction algorithm has been constructed and implemented
as \textsf{REDUCE} package \textsf{RECURSOR} in~\cite{GBGS:90,BGS:91,B:92}.
It also has been implemented as \textsf{FORM} package
\textsf{SHELL2}~\cite{FT:92}.

\begin{figure}[ht]
\begin{center}
\begin{picture}(42,17)
\put(21,8.5){\makebox(0,0){\includegraphics{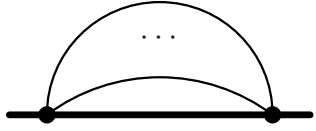}}}
\put(21,11){\makebox(0,0){{$\cdots$}}}
\end{picture}
\end{center}
\caption{$n$-loop on-shell sunset diagrams}
\label{F:Msun}
\end{figure}

\begin{figure}[ht]
\begin{center}
\begin{picture}(76,11)
\put(16,5.5){\makebox(0,0){\includegraphics{m1.eps}}}
\put(74,5.5){\makebox(0,0){\includegraphics{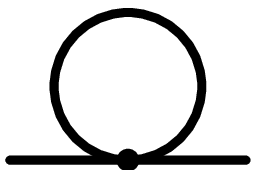}}}
\put(47,5.5){\makebox(0,0){{$\displaystyle{}=-\frac{1}{2}\frac{d-2}{d-3}$}}}
\end{picture}
\end{center}
\caption{1-loop on-shell diagram}
\label{F:Msun1}
\end{figure}

The first 2 basis integrals are trivial.
The $n$-loop on-shell sunset $M_n$ (Fig.~\ref{F:Msun}) is
\begin{gather}
M_n = \frac{(nd-4n+1)_{2(n-1)}}%
{\left(n+1-n\tfrac{d}{2}\right)_n
\left((n+1)\tfrac{d}{2}-2n-1\right)_n
\left(n\tfrac{d}{2}-2n+1\right)_{n-1}
\left(n-(n-1)\tfrac{d}{2}\right)_{n-1}}
\nonumber\\
\hphantom{M_n={}}
{}\times\frac{\Gamma(1+(n-1)\varepsilon)\Gamma(1+n\varepsilon)
\Gamma(1-2n\varepsilon)\Gamma^n(1-\varepsilon)}%
{\Gamma(1-n\varepsilon)\Gamma(1-(n+1)\varepsilon)}\,.
\label{M2:sun}
\end{gather}
In the 1-loop case, it can be reduced to the 1-loop massive vacuum diagram
\begin{equation}
M_1 = - \frac{1}{2} \frac{d-2}{d-3} V_1\,,\quad
V_1 = \frac{4\Gamma(1+\varepsilon)}{(d-2)(d-4)}
\label{M2:V}
\end{equation}
(Fig.~\ref{F:Msun1}).

Instead of using $N(1,1,1,0,0)$
(the last diagram in Fig.~\ref{F:M2b}, which is divergent)
as an element of the basis,
one may use the convergent diagram $N(1,1,1,1,1)$
(the second one in Fig.~\ref{F:M2t}).
It has been calculated in~\cite{B:92}:
\begin{equation}
\begin{split}
N(1,1,1,1,1) ={}&
\frac{4\Gamma^2(1+\varepsilon)}{3\varepsilon(1-4\varepsilon^2)}
\Biggl[ - 2\,_3 F_2 \left(
\begin{array}{c}
1,\frac{1}{2}-\varepsilon,\frac{1}{2}-\varepsilon\\
\frac{3}{2}+\varepsilon,\frac{3}{2}
\end{array}
\right|\left.\vphantom{\frac{1}{1}}1\right)\\
&{} + \frac{1}{1+2\varepsilon}\,_3 F_2 \left(
\begin{array}{c}
1,\frac{1}{2},\frac{1}{2}-\varepsilon\\
\frac{3}{2}+\varepsilon,\frac{3}{2}+\varepsilon
\end{array}
\right|\left.\vphantom{\frac{1}{1}}1\right)\\
&{} + \frac{3(1+2\varepsilon)}{16\varepsilon^2} \left(
\frac{\Gamma(1-4\varepsilon)\Gamma(1+2\varepsilon)\Gamma^2(1-\varepsilon)}%
{\Gamma(1-3\varepsilon)\Gamma(1-2\varepsilon)\Gamma(1+\varepsilon)}
- 1 \right) \Biggr]\,.
\end{split}
\label{M2:F}
\end{equation}
Its value at $\varepsilon=0$ is~\cite{B:90}
\begin{equation}
N(1,1,1,1,1) = \pi^2 \log2 - \frac{3}{2} \zeta_3
+ \mathcal{O}(\varepsilon)\,.
\label{M2:N1}
\end{equation}
D.~Broadhurst discovered~\cite{B:92} a symmetry group of a class of
${}_3 F_2$ functions like those in~(\ref{M2:F});
this allowed him to obtain~(\ref{M2:N1}) and two further terms,
up to $\mathcal{O}(\varepsilon^2)$, by purely algebraic methods.

\begin{figure}[ht]
\begin{center}
\begin{picture}(94,22)
\put(11,13){\makebox(0,0){\includegraphics{m2.eps}}}
\put(68,13){\makebox(0,0){\includegraphics{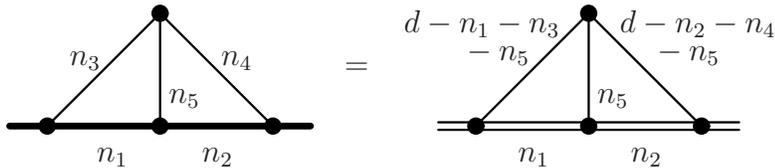}}}
\put(37,13){\makebox(0,0){{${}={}$}}}
\put(4.5,0){\makebox(0,0)[b]{{$n_1$}}}
\put(18.5,0){\makebox(0,0)[b]{{$n_2$}}}
\put(3,14){\makebox(0,0)[r]{{$n_3$}}}
\put(19,14){\makebox(0,0)[l]{{$n_4$}}}
\put(12,9){\makebox(0,0)[l]{{$n_5$}}}
\put(60.5,0){\makebox(0,0)[b]{{$n_1$}}}
\put(75.5,0){\makebox(0,0)[b]{{$n_2$}}}
\put(64,19){\makebox(0,0)[r]{{$d-n_1-n_3$}}}
\put(60,15){\makebox(0,0)[r]{{${}-n_5$}}}
\put(72,19){\makebox(0,0)[l]{{$d-n_2-n_4$}}}
\put(76,15){\makebox(0,0)[l]{{${}-n_5$}}}
\put(69,9){\makebox(0,0)[l]{{$n_5$}}}
\end{picture}
\end{center}
\caption{Inversion relation}
\label{F:M2i}
\end{figure}

Using inversion of Euclidean dimensionless momenta~(\ref{Q1:Inv})
and~(\ref{M1:Inv}), we obtain~\cite{BG:95a} relation of Fig.~\ref{F:M2i}.

\subsection{2 loops, 2 masses}
\label{S:M2m}

\begin{figure}[ht]
\begin{center}
\begin{picture}(42,13)
\put(21,6.5){\makebox(0,0){\includegraphics{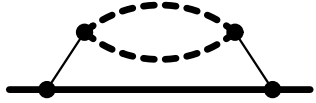}}}
\end{picture}
\end{center}
\caption{2-loop on-shell propagator diagram}
\label{F:M22}
\end{figure}

\begin{figure}[ht]
\begin{center}
\begin{picture}(32,17)
\put(16,8.5){\makebox(0,0){\includegraphics{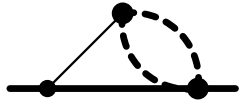}}}
\end{picture}
\end{center}
\caption{Topology of 2-loop on-shell propagator diagrams with 2 masses}
\label{F:M22t}
\end{figure}

In realistic theories, there are several massive particles with different masses,
as well as massless particles
(QED with $e$, $\mu$, $\tau$; QCD with $c$, $b$, $t$;
the Standard Model).
Therefore, diagrams like in Fig.~\ref{F:M22} appear.
Here the thin line is massless, the thick solid line has mass $m$,
and the thick dashed one has mass $m'$.
Combining identical denominators,
we obtain the topology shown in Fig.~\ref{F:M22t}.

\begin{figure}[ht]
\begin{center}
\begin{picture}(107,17)
\put(53.5,8.5){\makebox(0,0){\includegraphics{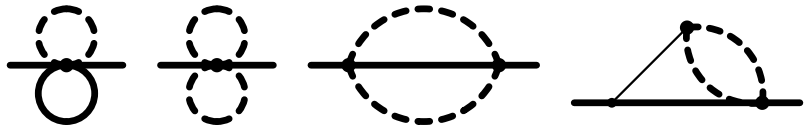}}}
\end{picture}
\end{center}
\caption{Basis 2-loop on-shell propagator diagrams with 2 masses}
\label{F:M22b}
\end{figure}

An algorithm for calculating such diagrams based on integration-by-parts
recurrence relations has been constructed in~\cite{DG:99}.
The algorithm has been implemented in \textsf{REDUCE}.
There is one scalar product which cannot be expressed via the denominators.
The integrals of Fig.~\ref{F:M22t} with arbitrary integer powers of the denominators,
and with arbitrary non-negative power of the scalar product in the numerator,
can be expressed as linear combinations of 4 basis integrals (Fig.~\ref{F:M22b});
coefficients are rational functions of $d$, $m$, and $m'$.

Instead of using the last diagram in Fig.~\ref{F:M22b} as an element of the basis,
one may prefer to use the sunset diagram (the third one in \ref{F:M22b})
with one of the denominators squared
(either the one with mass $m$ or the one with mass $m'$).
Therefore, the reduction algorithm for the sunset diagrams with the masses
$m$, $m'$, $m'$ and $p^2=m^2$, constructed in~\cite{OV:02},
is just a particular case of the algorithm from~\cite{DG:99}.

The first 2 basis integrals in Fig.~\ref{F:M22b} are just products
of 1-loop massive vacuum diagrams, and thus trivial.
The 2 nontrivial ones have been calculated in~\cite{DG:99}
using the Mellin-Barnes representation technique:
\begin{equation}
\begin{split}
\frac{I_0}{\Gamma^2(1+\varepsilon)} ={}&
- \frac{(m^{\prime2})^{1-2\varepsilon}}{\varepsilon^2(1-\varepsilon)}
\Biggl\{ \frac{1}{1-2\varepsilon}
{}_3F_2\left(
\begin{array}{c}
1,\frac{1}{2},-1+2\varepsilon\\
2-\varepsilon,\frac{1}{2}+\varepsilon
\end{array}
\right. \left|\frac{m^2}{m^{\prime2}}\right)\\
&{} + \left(\frac{m^2}{m^{\prime2}}\right)^{1-\varepsilon}
{}_3F_2\left(
\begin{array}{c}
1,\varepsilon,\frac{3}{2}-\varepsilon\\
3-2\varepsilon,\frac{3}{2}
\end{array}
\right. \left|\frac{m^2}{m^{\prime2}}\right)
\Biggr\}\,,\\
\frac{I_1}{\Gamma^2(1+\varepsilon)} ={}&
- \frac{(m^{\prime2})^{-2\varepsilon}}{\varepsilon^2}
\Biggl\{ \frac{1}{2(1-\varepsilon)(1+2\varepsilon)}
{}_3F_2\left(
\begin{array}{c}
1,\frac{1}{2},2\varepsilon\\
2-\varepsilon,\frac{3}{2}+\varepsilon
\end{array}
\right. \left|\frac{m^2}{m^{\prime2}}\right)\\
&{} - \left(\frac{m^2}{m^{\prime2}}\right)^{1-\varepsilon}
\frac{1}{1-2\varepsilon}
{}_3F_2\left(
\begin{array}{c}
1,\varepsilon,\frac{1}{2}-\varepsilon\\
2-2\varepsilon,\frac{3}{2}
\end{array}
\right. \left|\frac{m^2}{m^{\prime2}}\right)
\Biggr\}\,.
\end{split}
\label{M2m:I}
\end{equation}

If $m'=m$, they are not independent -- $I_1$ can be reduced to $I_0$
(the last diagram in Fig.~\ref{F:M2b}) using integration by parts (Sect.~\ref{S:M2}):
\begin{equation}
I_1 = - \frac{3d-8}{4(d-4)} I_0 - \frac{3(d-2)^2}{8(d-3)(d-4)} V_1^2\,.
\label{M2m:II}
\end{equation}
We can express $N(1,1,1,1,1)$ via $I_0$ or $I_1$~(\ref{M2m:I});
in the second case, for example, we have
\begin{equation}
\begin{split}
N(1,1,1,1,1) ={}& - \frac{\Gamma^2(1+\varepsilon)}{(d-4)^2}
\Biggl\{ (3d-10) \biggl[ \frac{1}{(d-2)(d-5)}
{}_3 F_2 \left(
\begin{array}{c}
1,2\varepsilon,\frac{1}{2}\\
2-\varepsilon,\frac{3}{2}+\varepsilon
\end{array}
\right|\left.\vphantom{\frac{1}{1}}1\right)\\
&{} + \frac{1}{d-3}
{}_3 F_2 \left(
\begin{array}{c}
1,\varepsilon,\frac{1}{2}-\varepsilon\\
2-2\varepsilon,\frac{3}{2}
\end{array}
\right|\left.\vphantom{\frac{1}{1}}1\right)
\biggr]\\
&{} + \frac{1}{(d-3)(d-4)} \left[ 2
\frac{\Gamma^2(1-\varepsilon)\Gamma(1+2\varepsilon)\Gamma(1-4\varepsilon)}%
{\Gamma(1+\varepsilon)\Gamma(1-2\varepsilon)\Gamma(1-3\varepsilon)}
- 3 \right] \Biggr\}\,.
\end{split}
\label{M2m:N1}
\end{equation}
This should be equivalent to~(\ref{M2:F}).

The basis integrals~(\ref{M2m:I}),
expanded~\cite{DG:99} in $\varepsilon$ up to $\mathcal{O}(1)$,
\begin{equation}
\begin{split}
\frac{I_0}{\Gamma^2(1+\varepsilon)} ={}&
- m^{2-4\varepsilon}
\left[ \frac{1}{2\varepsilon^2} + \frac{5}{4\varepsilon}
+ 2 (1-r^2)^2 (L_+ + L_-) - 2 \log^2 r + \frac{11}{8} \right]\\
&{} - m^{\prime2-4\varepsilon}
\left[ \frac{1}{\varepsilon^2} + \frac{3}{\varepsilon}
- 2 \log r + 6 \right]
+ \mathcal{O}(\varepsilon)\,,\\
\frac{I_1}{\Gamma^2(1+\varepsilon)} ={}&
m^{-4\varepsilon}
\left[ \frac{1}{2\varepsilon^2} + \frac{5}{2\varepsilon}
+ 2 (1+r)^2 L_+ + 2 (1-r)^2 L_-
- 2 \log^2 r + \frac{19}{2} \right]\\
&{} + \mathcal{O}(\varepsilon)\,,
\end{split}
\label{M2m:Ie}
\end{equation}
(where $r = m'/m$),
are expressed via dilogarithms:
\begin{equation}
\begin{split}
L_+ &{}= - \Li2(-r) + \tfrac{1}{2} \log^2 r - \log r \log(1+r) - \tfrac{1}{6} \pi^2\\
&{}= \Li2(-r^{-1}) + \log r^{-1} \log(1+r^{-1})\,,\\
L_- &{}= \Li2(1-r) + \tfrac{1}{2} \log^2 r + \tfrac{1}{6} \pi^2\\
&{}= - \Li2(1-r^{-1}) + \tfrac{1}{6} \pi^2\\
&{}= - \Li2(r) + \tfrac{1}{2} \log^2 r - \log r \log(1-r) + \tfrac{1}{3} \pi^2 \qquad (r<1)\\
&{}= \Li2(r^{-1}) + \log r^{-1} \log(1-r^{-1}) \qquad (r>1)\,,\\
L_+ + L_- &{}= \tfrac{1}{2} \Li2(1-r^2) + \log^2 r + \tfrac{1}{12} \pi^2\\
&{}= - \tfrac{1}{2} \Li2(1-r^{-2}) + \tfrac{1}{12} \pi^2\,.
\end{split}
\label{QCDos:Lpm}
\end{equation}
The sunset basis integral $I_0$~(\ref{M2m:I}) (the third one in Fig.~\ref{F:M22b})
was recently expanded~\cite{AMR:02} up to $\varepsilon^5$;
the coefficients involve harmonic polylogarithms of $r^{-1}$.
In particular, at $r=1$, one more term in~(\ref{M2:N1})
($\mathcal{O}(\varepsilon^3)$) can be obtained.

\subsection{3 loops}
\label{S:M3}

\begin{figure}[p]
\begin{center}
\begin{picture}(132,109)
\put(66,54.5){\makebox(0,0){\includegraphics{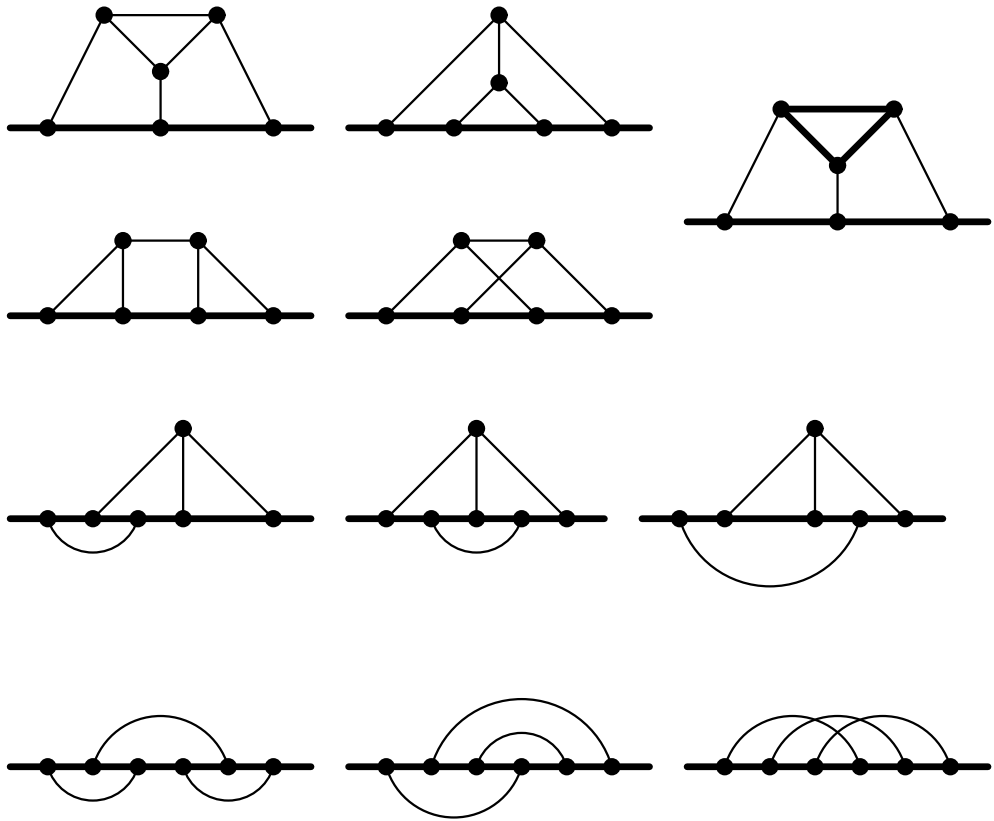}}}
\end{picture}
\end{center}
\caption{Topologies of 3-loop massive on-shell propagator diagrams}
\label{F:M3t}
\end{figure}

\begin{figure}[p]
\begin{center}
\begin{picture}(160,57)
\put(80,28.5){\makebox(0,0){\includegraphics[width=160mm]{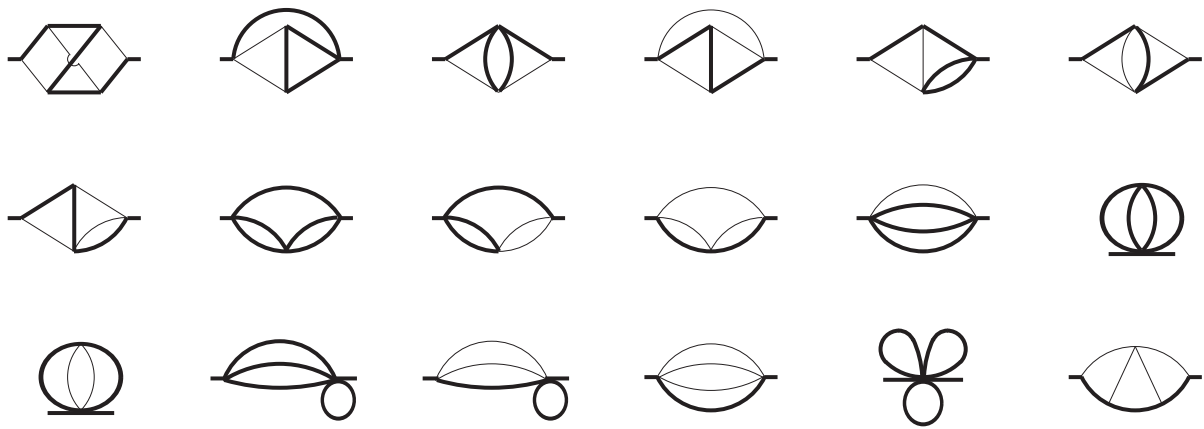}}}
\end{picture}
\end{center}
\caption{Basis diagrams (all indices equal 1, no numerators)}
\label{F:M3b}
\end{figure}

There are 11 generic topologies of 3-loop massive on-shell propagator diagrams
(Fig.~\ref{F:M3t}).
Ten of them are the same as in HQET (Fig.~\ref{F:H3t}),
plus one additional topology with a heavy loop
(in HQET, diagrams with heavy-quark loops vanish).
Integration-by-parts recurrence relations for these diagrams
have been investigated in~\cite{MR:00c}.
They can be used to reduce all integrals of Fig.~\ref{F:M3t},
with arbitrary integer powers of denominators
and arbitrary numerators,
to linear combinations of the basis integrals in Fig.~\ref{F:M3b}.
This algorithm has been implemented in the \textsf{FORM}
package \textsf{SHELL3}~\cite{MR:00c}.
Some of the basis integrals are trivial.
Others were found, to some orders in $\varepsilon$,
during many years, in the course of calculation
of the anomalous magnetic moment in QED at 3 loops~\cite{LR:96}.
Some additional terms of $\varepsilon$ expansions
were obtained in~\cite{MR:00c}.

\begin{figure}[ht]
\begin{center}
\begin{picture}(92,50)
\put(36,25){\makebox(0,0){\includegraphics{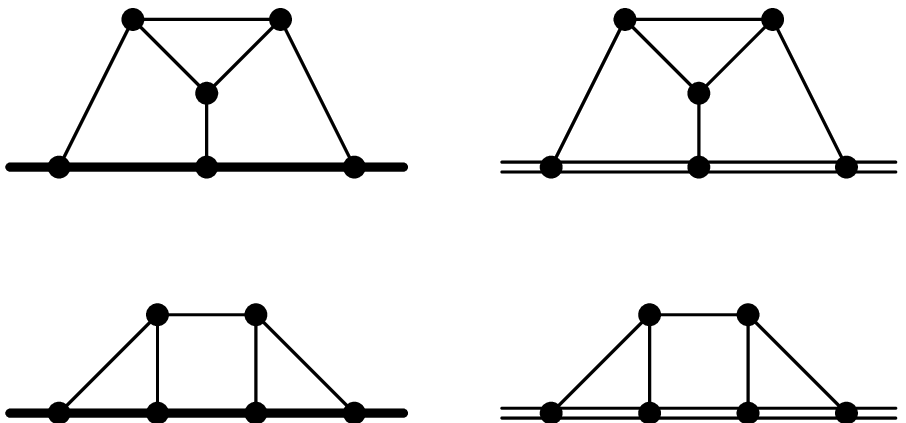}}}
\put(3.5,26){\makebox(0,0)[b]{$n_1$}}
\put(18.5,26){\makebox(0,0)[b]{$n_2$}}
\put(53.5,26){\makebox(0,0)[b]{$n_1$}}
\put(68.5,26){\makebox(0,0)[b]{$n_2$}}
\put(-1.5,37.5){\makebox(0,0)[r]{$n_3$}}
\put(23,37.5){\makebox(0,0)[l]{$n_4$}}
\put(51,43){\makebox(0,0)[r]{$d-n_1-n_3$}}
\put(49,39){\makebox(0,0)[r]{${}-n_5-n_7$}}
\put(70.5,43){\makebox(0,0)[l]{$d-n_2-n_4$}}
\put(72,39){\makebox(0,0)[l]{${}-n_5-n_8$}}
\put(12,33.75){\makebox(0,0)[l]{$n_5$}}
\put(62,33.75){\makebox(0,0)[l]{$n_5$}}
\put(7.7,39){\makebox(0,0)[r]{$n_7$}}
\put(15,39){\makebox(0,0)[l]{$n_8$}}
\put(57.7,39){\makebox(0,0)[r]{$n_7$}}
\put(65,39){\makebox(0,0)[l]{$n_8$}}
\put(11,46){\makebox(0,0)[b]{$n_6$}}
\put(61,46){\makebox(0,0)[b]{$d-n_6-n_7-n_8$}}
\put(36,30){\makebox(0,0){$=$}}
\put(1,1){\makebox(0,0)[b]{$n_1$}}
\put(11,1){\makebox(0,0)[b]{$n_3$}}
\put(21,1){\makebox(0,0)[b]{$n_2$}}
\put(51,1){\makebox(0,0)[b]{$n_1$}}
\put(61,1){\makebox(0,0)[b]{$n_3$}}
\put(71,1){\makebox(0,0)[b]{$n_2$}}
\put(-0.5,10){\makebox(0,0)[r]{$n_4$}}
\put(22,10){\makebox(0,0)[l]{$n_5$}}
\put(51.5,14){\makebox(0,0)[r]{$d-n_1-n_4$}}
\put(47.5,9.5){\makebox(0,0)[r]{${}-n_6$}}
\put(70,14){\makebox(0,0)[l]{$d-n_2-n_5$}}
\put(73,9.5){\makebox(0,0)[l]{${}-n_7$}}
\put(6.2,10){\makebox(0,0)[l]{$n_6$}}
\put(15.8,10){\makebox(0,0)[r]{$n_7$}}
\put(56.2,10){\makebox(0,0)[l]{$n_6$}}
\put(65.8,10){\makebox(0,0)[r]{$n_7$}}
\put(11,16){\makebox(0,0)[b]{$n_8$}}
\put(61,16){\makebox(0,0)[b]{$d-n_3-n_6-n_7-n_8$}}
\put(36,5){\makebox(0,0){$=$}}
\end{picture}
\end{center}
\caption{Inversion relations}
\label{F:M3i}
\end{figure}

\begin{figure}[ht]
\begin{center}
\begin{picture}(100,13)
\put(1,0){\makebox(0,0)[b]{\includegraphics{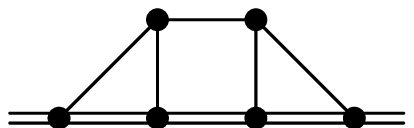}}}
\put(48,0){\makebox(0,0)[b]{\includegraphics{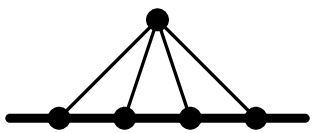}}}
\put(27,4){\makebox(0,0){{${}={}$}}}
\put(65,4){\makebox(0,0)[l]{{${}=-5\zeta_5+12\zeta_2\zeta_3$}}}
\end{picture}
\end{center}
\caption{Inversion relation: all $n_i=1$, $d=4$}
\label{F:M3i2}
\end{figure}

Performing inversion~(\ref{Q1:Inv}) of the loop momenta,
we obtain the relations in Fig.~\ref{F:M3i}.
For example, the HQET ladder diagram with all indices $n_i=1$ is convergent;
its value at $d=4$ is related~\cite{CM:02}
to a massive on-shell diagram (Fig.~\ref{F:M3i2})
by the second inversion relation.
This is one of the basis diagrams of Fig.~\ref{F:M3b},
and its value at $d=4$ is known (Fig.~\ref{F:M3i2}).
Calculating this ladder diagram with \textsf{Grinder} (Sect.~\ref{S:H3})
and solving for the most difficult HQET 3-loop basis integral in Fig.~\ref{F:H3b},
we obtain the $\varepsilon$ expansion of this integral
up to $\mathcal{O}(\varepsilon)$.
This concludes the investigation of the basis integrals of Fig.~\ref{F:H3b}
(Sects.~\ref{S:H3}--\ref{S:In}),
and allows one to solve 3-loop propagator problems in HQET
up to terms $\mathcal{O}(1)$ (see~\cite{CG:03}).

\FloatBarrier
\section{Hypergeometric functions and multiple $\zeta$ values}
\label{S:hyper}

In this Section, we shall consider some mathematical methods
useful for expanding massless and HQET diagrams in $\varepsilon$.

\subsection{Multiple $\zeta$ values}
\label{S:zeta}

As you may have noticed, expansions of various Feynman integrals in $\varepsilon$
contain lots of $\zeta$-functions.
They appear when expanding $\Gamma$-functions:
\begin{equation}
\Gamma(1+\varepsilon) = \exp \left[ - \gamma\varepsilon
+ \sum_{n=2}^\infty \frac{(-1)^n \zeta_n}{n} \varepsilon^n \right]\,.
\label{zeta:Gamma}
\end{equation}
They also appear, together with their generalizations -- multiple $\zeta$ values,
when expanding hypergeometric functions of unit argument
with indices going to integers at $\varepsilon\to0$
(Sect.~\ref{S:exam}, \ref{S:algo}).
Such hypergeometric functions appear in massless and HQET diagrams
(Sect.~\ref{S:Gn}, \ref{S:Jn}, \ref{S:In}).
In this Section, we shall briefly discuss multiple $\zeta$ values;
more details can be found in~\cite{BBBL:01},
where more general functions are also considered.

The Riemann $\zeta$-function is defined by
\begin{equation}
\zeta_s = \sum_{n>0} \frac{1}{n^s}\,.
\label{zeta:zeta1}
\end{equation}
Let's define multiple $\zeta$ values as
\begin{equation}
\begin{split}
&\zeta_{s_1 s_2} = \sum_{n_1>n_2>0} \frac{1}{n_1^{s_1}n_2^{s_2}}\,,\\
&\zeta_{s_1 s_2 s_3} = \sum_{n_1>n_2>n_3>0}
\frac{1}{n_1^{s_1}n_2^{s_2}n_3^{s_3}}\,,
\end{split}
\label{zeta:zeta2}
\end{equation}
and so on.
These series converge at $s_1>1$.
The number of summations is called \emph{depth} $k$;
$s=s_1+\cdots+s_k$ is called \emph{weight}.
All identities we shall consider relate terms of the same weight
(where the weight of a product is the sum of the factors' weights).

For example, there is 1 convergent sum at weight 2:
\begin{equation*}
\zeta_2\,,
\end{equation*}
2 at weight 3:
\begin{equation*}
\zeta_{3}\,,\quad
\zeta_{21}\,,
\end{equation*}
4 at weight 4:
\begin{equation*}
\zeta_4\,,\quad
\zeta_{31}\,,\quad
\zeta_{22}\,,\quad
\zeta_{211}\,,
\end{equation*}
and 8 at weight 5:
\begin{equation*}
\zeta_5\,,\quad
\zeta_{41}\,,\quad
\zeta_{32}\,,\quad
\zeta_{23}\,,\quad
\zeta_{311}\,,\quad
\zeta_{221}\,,\quad
\zeta_{212}\,,\quad
\zeta_{2111}\,.
\end{equation*}

Suppose we want to multiply $\zeta_s \zeta_{s_1 s_2}$:
\begin{equation*}
\zeta_s \zeta_{s_1 s_2} =
\sum_{\substack{n>0\\n_1>n_2>0}} \frac{1}{n^s n_1^{s_1} n_2^{s_2}}\,.
\end{equation*}
Here $n$ can be anywhere with respect to $n_1$, $n_2$.
There are 5 contributions:
\begin{gather*}
\sum_{n>n_1>n_2>0} \frac{1}{n^s n_1^{s_1} n_2^{s_2}}
= \zeta_{s s_1 s_2}\,,\\
\sum_{n=n_1>n_2>0} \frac{1}{n^s n_1^{s_1} n_2^{s_2}}
= \zeta_{s+s_1,s_2}\,,\\
\sum_{n_1>n>n_2>0} \frac{1}{n^s n_1^{s_1} n_2^{s_2}}
= \zeta_{s_1 s s_2}\,,\\
\sum_{n_1>n=n_2>0} \frac{1}{n^s n_1^{s_1} n_2^{s_2}}
= \zeta_{s_1,s+s_2}\,,\\
\sum_{n_1>n_2>n>0} \frac{1}{n^s n_1^{s_1} n_2^{s_2}}
= \zeta_{s_1 s_2 s}\,.
\end{gather*}
Therefore,
\begin{equation}
\zeta_s \zeta_{s_1 s_2}
= \zeta_{s s_1 s_2}
+ \zeta_{s+s_1,s_2}
+ \zeta_{s_1 s s_2}
+ \zeta_{s_1,s+s_2}
+ \zeta_{s_1 s_2 s}\,.
\label{zeta:stuffle}
\end{equation}

\begin{figure}[ht]
\begin{center}
\begin{picture}(163,122)
\put(81.5,61){\makebox(0,0){\includegraphics{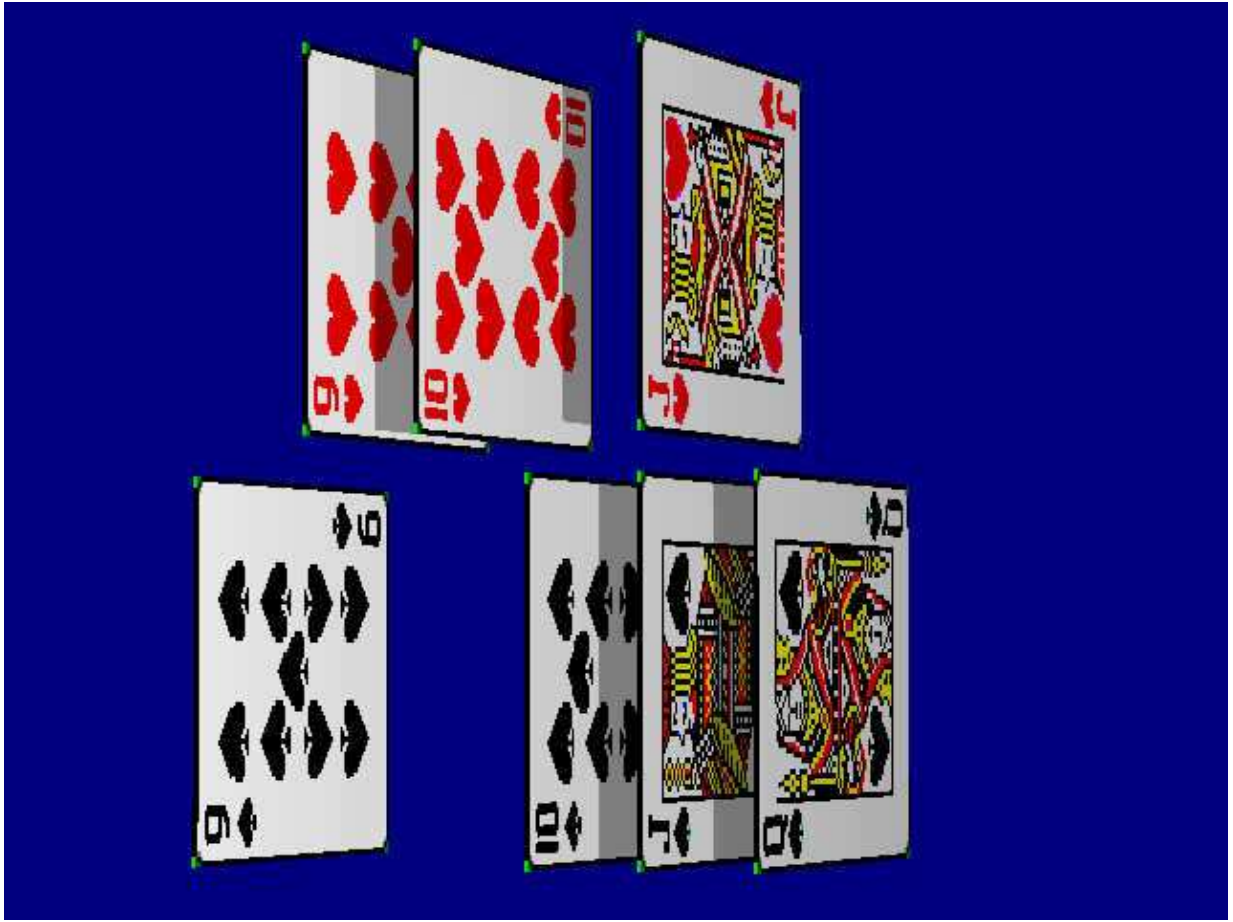}}}
\end{picture}
\end{center}
\caption{Stuffling: shuffling with (possible) stuffing}
\label{F:Stuffling}
\end{figure}

This process reminds shuffling cards (Fig.~\ref{F:Stuffling}).
The order of cards in the upper deck, as well as in the lower one,
is kept fixed.
We sum over all possible shufflings.
Unlike real playing cards, however,
two cards may be exactly on top of each other.
In this case, they are stuffed together:
a single card (which is their sum) appears in the resulting deck.
A mathematical jargon term for such shuffling with (possible) stuffing
is \emph{stuffling}, see~\cite{BBBL:01}.

For example, the stuffling relations for weights $s\le5$ are
\begin{equation}
\begin{split}
\zeta_2^2 &{}= \zeta_{22} + \zeta_4 + \zeta_{22}\,,\\
\zeta_2 \zeta_3 &{}= \zeta_{23} + \zeta_5 + \zeta_{32}\,,\\
\zeta_2 \zeta_{21} &{}=
\zeta_{221} + \zeta_{41} + \zeta_{221} + \zeta_{23} + \zeta_{212}\,.
\end{split}
\label{zeta:stuffle5}
\end{equation}

Now we are going to derive an integral representation
of multiple $\zeta$ values.
Let's consider the integral
\begin{gather*}
\int_0^1 \frac{dx_1}{x_1}
\int_0^{x_1} \frac{dx_2}{x_2}
\int_0^{x_2} \frac{dx_3}{x_3}
\int_0^{x_3} \frac{dx_4}{x_4}
x_4^n ={}\\
\int_0^1 \frac{dx_1}{x_1}
\int_0^{x_1} \frac{dx_2}{x_2}
\int_0^{x_2} \frac{dx_3}{x_3}
x_3^n
\cdot \frac{1}{n} ={}\\
\int_0^1 \frac{dx_1}{x_1}
\int_0^{x_1} \frac{dx_2}{x_2}
x_2^n
\cdot \frac{1}{n^2} ={}\\
\int_0^1 \frac{dx_1}{x_1}
x_1^n
\cdot \frac{1}{n^3} ={}\\
\frac{1}{n^4}\,.
\end{gather*}
It is easy to guess that
\begin{equation*}
\frac{1}{n^s} =
\int_{1>x_1>\cdots>x_s>0}
\frac{dx_1}{x_1} \cdots \frac{dx_s}{x_s} x_s^n\,.
\end{equation*}
But we need the sum~(\ref{zeta:zeta1}).
This is also easy:
\begin{equation}
\zeta_s =
\int_{1>x_1>\cdots>x_s>0}
\frac{dx_1}{x_1} \cdots \frac{dx_{s-1}}{x_{s-1}} \frac{dx_s}{1-x_s}\,.
\label{zeta:int}
\end{equation}

Let's introduce short notations:
\begin{equation*}
\omega_0 = \frac{dx}{x}\,,\quad
\omega_1 = \frac{dx}{1-x}\,.
\end{equation*}
All integrals will always have the integration region $1>x_1>\cdots>x_s>0$.
Then~(\ref{zeta:int}) becomes
\begin{equation}
\zeta_s = \int \omega_0^{s-1} \omega_1\,.
\label{zeta:int1}
\end{equation}
This can be generalized to multiple sums:
\begin{equation}
\begin{split}
\zeta_{s_1 s_2} &{}= \int \omega_0^{s_1-1} \omega_1\; \omega_0^{s_2-1} \omega_1\,,\\
\zeta_{s_1 s_2 s_3} &{}=
\int \omega_0^{s_1-1} \omega_1\; \omega_0^{s_2-1} \omega_1\; \omega_0^{s_3-1} \omega_1\,,
\end{split}
\label{zeta:int2}
\end{equation}
and so on.

Suppose we want to multiply $\zeta_2 \zeta_2$:
\begin{equation*}
\zeta_2^2 =
\int_{1>x_1>x_2>0}\omega_0\omega_1 \cdot
\int_{1>x_1'>x_2'>0}\omega_0\omega_1\,.
\end{equation*}
The ordering of primed and non-primed integration variables is not fixed.
There are 6 contributions:
\begin{gather*}
1>x_1>x_2>x_1'>x_2'>0:\quad
\int \omega_0 \omega_1 \omega_0 \omega_1 = \zeta_{22}\,,\\
1>x_1>x_1'>x_2>x_2'>0:\quad
\int \omega_0 \omega_0 \omega_1 \omega_1 = \zeta_{31}\,,\\
1>x_1>x_1'>x_2'>x_2>0:\quad
\int \omega_0 \omega_0 \omega_1 \omega_1 = \zeta_{31}\,,\\
1>x_1'>x_1>x_2>x_2'>0:\quad
\int \omega_0 \omega_0 \omega_1 \omega_1 = \zeta_{31}\,,\\
1>x_1'>x_1>x_2'>x_2>0:\quad
\int \omega_0 \omega_0 \omega_1 \omega_1 = \zeta_{31}\,,\\
1>x_1'>x_2'>x_1>x_2>0:\quad
\int \omega_0 \omega_1 \omega_0 \omega_1 = \zeta_{22}\,.
\end{gather*}
Therefore,
\begin{equation}
\zeta_2^2 = 4 \zeta_{31} + 2 \zeta_{22}\,.
\label{zeta:shuffling2}
\end{equation}

\begin{figure}[ht]
\begin{center}
\begin{picture}(163,122)
\put(81.5,61){\makebox(0,0){\includegraphics{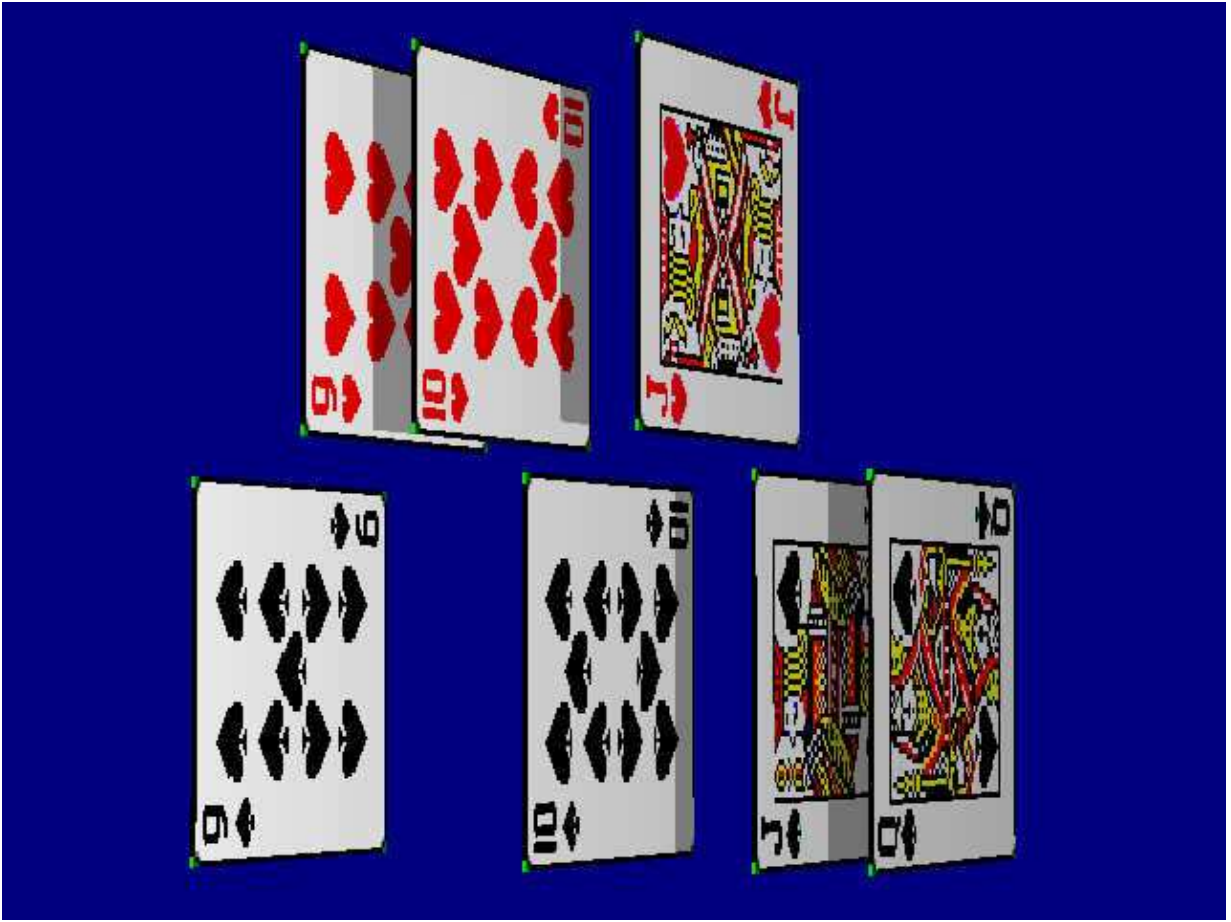}}}
\end{picture}
\end{center}
\caption{Shuffling}
\label{F:Shuffling}
\end{figure}

Now we are multiplying integrals, not sums.
Therefore, our cards are now infinitely thin,
and cannot be exactly on top of each other.
There are just 2 kinds of cards: $\omega_0$ and $\omega_1$,
and we sum over all possible shufflings of two decks
(Fig.~\ref{F:Shuffling}).

For example, the shuffling relations for weight 5 are
\begin{equation}
\begin{split}
&\zeta_2 \zeta_3 = 6 \zeta_{311} + 3 \zeta_{221} + \zeta_{212}\,,\\
&\zeta_2 \zeta_{21} = \zeta_{41} + 2 \zeta_{32} + 2 \zeta_{23}\,.
\end{split}
\label{zeta:shuffling3}
\end{equation}

The integral representation allows us to derive
another set of useful relations, even simpler than shuffling --
duality relations.
Let's make the substitution $x_i\to1-x_i$.
Then $\omega_0\leftrightarrow\omega_1$;
in order to preserve the ordering $1>x_1>\cdots x_s>0$,
we have to arrange all the $\omega$ factors in the opposite order.
In other words, after writing down an integral representation
for a multiple $\zeta$ value,
we may read it in the Arabic fashion, from right to left,
simultaneously replacing $\omega_0\leftrightarrow\omega_1$:
\begin{equation}
\begin{split}
&\zeta_3 = \int\omega_0\omega_0\omega_1
= \int\omega_0\omega_1\omega_1 = \zeta_{21}\,,\\
&\zeta_4 = \int\omega_0\omega_0\omega_0\omega_1
= \int\omega_0\omega_1\omega_1\omega_1 = \zeta_{211}\,,\\
&\zeta_5 = \int\omega_0\omega_0\omega_0\omega_0\omega_1
= \int\omega_0\omega_1\omega_1\omega_1\omega_1 = \zeta_{2111}\,,\\
&\zeta_{41} = \int\omega_0\omega_0\omega_0\omega_1\omega_1
= \int\omega_0\omega_0\omega_1\omega_1\omega_1 = \zeta_{311}\,,\\
&\zeta_{32} = \int\omega_0\omega_0\omega_1\omega_0\omega_1
= \int\omega_0\omega_1\omega_0\omega_1\omega_1 = \zeta_{221}\,,\\
&\zeta_{23} = \int\omega_0\omega_1\omega_0\omega_0\omega_1
= \int\omega_0\omega_1\omega_1\omega_0\omega_1 = \zeta_{212}\,.
\end{split}
\label{zeta:duality}
\end{equation}
Duality relations are the only known relations which say that two distinct
multiple $\zeta$ values are just equal to each other.

Let's summarize.
There are 3 distinct multiple $\zeta$ values of weight 4,
\begin{equation*}
\zeta_4 = \zeta_{211}\,,\quad
\zeta_{31}\,,\quad
\zeta_{22}\,,
\end{equation*}
due to duality~(\ref{zeta:duality}).
The stuffling~(\ref{zeta:stuffle5}) and shuffling~(\ref{zeta:shuffling2})
relations yield
\begin{equation*}
\begin{array}{lcl}
\zeta_2^2 = \zeta_4 + 2 \zeta_{22}      & \to &
\displaystyle \zeta_{22} = \frac{3}{4} \zeta_4\,,\\[2mm]
\zeta_2^2 = 4 \zeta_{31} + 2 \zeta_{22} & \to &
\displaystyle \zeta_{31} = \frac{1}{4} \zeta_4\,,
\end{array}
\end{equation*}
where we have used
\begin{equation*}
\zeta_2^2 = \frac{5}{2} \zeta_4\,.
\end{equation*}
This follows from
\begin{equation}
\zeta_2 = \frac{\pi^2}{6}\,,\quad
\zeta_4 = \frac{\pi^4}{90}\,.
\label{zeta:24}
\end{equation}
Therefore, all multiple $\zeta$ values of weight 4
can be expressed via $\zeta_4$.

There are 4 distinct multiple $\zeta$ values of weight 5,
\begin{equation*}
\zeta_5 = \zeta_{2111}\,,\quad
\zeta_{41} = \zeta_{311}\,,\quad
\zeta_{32} = \zeta_{221}\,,\quad
\zeta_{23} = \zeta_{212}\,,
\end{equation*}
due to duality~(\ref{zeta:duality}).
The stuffling~(\ref{zeta:stuffle5}) and shuffling~(\ref{zeta:shuffling3})
relations yield
\begin{equation*}
\left\{
\begin{array}{l}
\zeta_2 \zeta_3 = \zeta_{32} + \zeta_{23} + \zeta_5\,,\\
\zeta_2 \zeta_3 = 6 \zeta_{41} + 3 \zeta_{32} + \zeta_{23}\,,\\
\zeta_2 \zeta_3 = \zeta_2 \zeta_{21} = \zeta_{41} + 2 \zeta_{32} + 2 \zeta_{23}\,.
\end{array}
\right.
\end{equation*}
Solving this linear system, we obtain
\begin{equation}
\zeta_{41} = 2 \zeta_5 - \zeta_2 \zeta_3\,,\quad
\zeta_{32} = - \frac{11}{2} \zeta_5 + 3 \zeta_2 \zeta_3\,,\quad
\zeta_{23} = \frac{9}{2} \zeta_5 - 2 \zeta_2 \zeta_3\,.
\label{zeta:weight5}
\end{equation}
Therefore, all multiple $\zeta$ values of weight 5
can be expressed via $\zeta_5$ and $\zeta_2\zeta_3$.

Weight 5 is sufficient for calculating finite parts of 3-loop propagator diagrams.
We shall not discuss higher weights here.

\subsection{Expanding hypergeometric functions in $\varepsilon$: example}
\label{S:exam}

There is an algorithm to expand in $\varepsilon$
hypergeometric functions of unit argument
whose indices tend to integers at $\varepsilon\to0$.
The coefficients are linear combinations of multiple $\zeta$ values.
We shall first consider an example:
expanding the ${}_3 F_2$ function~(\ref{Gn:Fe}),
which appears in $G(1,1,1,1,2+\varepsilon)$ (Sect.~\ref{S:Gn}),
up to $\mathcal{O}(\varepsilon)$.

The function we want to expand is
\begin{equation}
F = \sum_{n=0}^\infty
\frac{(2-2\varepsilon)_n(2+2\varepsilon)_n}{(3+\varepsilon)_n(3+2\varepsilon)_n}\,.
\label{exam:F}
\end{equation}
When arguments of two Pochhammer symbols differ by an integer,
their ratio can be simplified:
\begin{equation*}
\frac{(2+2\varepsilon)_n}{(3+2\varepsilon)_n} =
\frac{2+2\varepsilon}{n+2+2\varepsilon}\,.
\end{equation*}

\textbf{Step 1}. We rewrite all Pochhammer symbols $(m+l\varepsilon)_n$
via $(1+l\varepsilon)_{n+m-1}$:
\begin{equation*}
(2-2\varepsilon)_n = \frac{(1-2\varepsilon)_{n+1}}{1-2\varepsilon}\,,\quad
(3+\varepsilon)_n = \frac{(1+\varepsilon)_{n+2}}{(1+\varepsilon)(2+\varepsilon)}\,.
\end{equation*}
Then our function $F$~(\ref{exam:F}) becomes
\begin{equation}
F = \frac{2(2+\varepsilon)(1+\varepsilon)^2}{1-2\varepsilon} F' =
2 (2+9\varepsilon+\cdots) F'\,,\quad
F' = \sum_{n=0}^\infty
\frac{1}{n+2+2\varepsilon}
\frac{(1-2\varepsilon)_{n+1}}{(1+\varepsilon)_{n+2}}\,.
\label{exam:F1}
\end{equation}
In what follows, we shall consider $F'$, and return to $F$
at the end of calculation.
It is convenient to introduce the function $P_n(\varepsilon)$:
\begin{equation}
(1+\varepsilon)_n = n!\,P_n(\varepsilon)\,,\quad
P_n(\varepsilon) = \prod_{n'=1}^n \left(1+\frac{\varepsilon}{n'}\right)\,.
\label{exam:P}
\end{equation}
Then
\begin{equation*}
F' = \sum_{n=0}^\infty
\frac{1}{(n+2)(n+2+2\varepsilon)}
\frac{P_{n+1}(-2\varepsilon)}{P_{n+2}(\varepsilon)}\,.
\end{equation*}

\textbf{Step 2}. We expand the rational function in front of $P$'s
in $\varepsilon$:
\begin{equation*}
\frac{1}{(n+2)(n+2+2\varepsilon)} =
\frac{1}{(n+2)^2} - \frac{2\varepsilon}{(n+2)^3} + \cdots
\end{equation*}
In this simple case, all $n$-dependent brackets are the same ($n+2$);
in more general cases, we should have to decompose each term into partial fractions
with respect to $n$.
Then
\begin{equation*}
F' = \sum_{n=0}^\infty \frac{1}{(n+2)^2}
\frac{P_{n+1}(-2\varepsilon)}{P_{n+2}(\varepsilon)}
- 2 \varepsilon \sum_{n=0}^\infty \frac{1}{(n+2)^3}
+ \mathcal{O}(\varepsilon^2)\,,
\end{equation*}
because we may replace all $P_n(l\varepsilon)\to1$
in the $\mathcal{O}(\varepsilon)$ term.
Shifting the summation indices, we get
\begin{equation*}
F' = \sum_{n=2}^\infty \frac{1}{n^2}
\frac{P_{n-1}(-2\varepsilon)}{P_n(\varepsilon)}
- 2\varepsilon \sum_{n=2}^\infty \frac{1}{n^3}
+ \mathcal{O}(\varepsilon^2)\,.
\end{equation*}

\textbf{Step 3}. We rewrite all $P_{n+m}(l\varepsilon)$ via $P_{n-1}(l\varepsilon)$:
\begin{equation*}
P_n(\varepsilon) = \left(1+\frac{\varepsilon}{n}\right) P_{n-1}(\varepsilon)\,.
\end{equation*}
We again expand in $\varepsilon$ (and, if necessary, decompose into partial fractions)
rational functions in front of $P$'s:
\begin{equation*}
F' = \sum_{n=2}^\infty \frac{1}{n^2}
\frac{P_{n-1}(-2\varepsilon)}{P_{n-1}(\varepsilon)}
- 3 \varepsilon \sum_{n=2}^\infty \frac{1}{n^3}
+ \mathcal{O}(\varepsilon^2)\,.
\end{equation*}

\textbf{Step 4}. We add and subtract a few first terms to have all summations
start from $n=1$:
\begin{equation*}
F' = \sum_{n=1}^\infty \frac{1}{n^2}
\frac{P_{n-1}(-2\varepsilon)}{P_{n-1}(\varepsilon)}
- 1 - 3 (\zeta_3-1) \varepsilon
+ \mathcal{O}(\varepsilon^2)\,.
\end{equation*}

\textbf{Step 5}. We expand $P_{n-1}(l\varepsilon)$ in $\varepsilon$:
\begin{equation}
P_{n-1}(\varepsilon) = \prod_{n>n'>0}
\left(1+\frac{\varepsilon}{n'}\right) =
1 + z_1(n) \varepsilon + z_{11}(n) \varepsilon^2 + \cdots
\label{exam:Pe}
\end{equation}
where
\begin{equation}
\begin{split}
&z_{s}(n) = \sum_{n>n_1>0} \frac{1}{n_1^s}\,,\\
&z_{s_1 s_2}(n) = \sum_{n>n_1>n_2>0} \frac{1}{n_1^{s_1} n_2^{s_2}}\,,
\end{split}
\label{exam:z}
\end{equation}
and so on.
These finite $z$ sums obey the same stuffling relations
as infinite $\zeta$ sums, for example,
\begin{equation}
z_s(n) z_{s_1 s_2}(n) = z_{s s_1 s_2}(n) + z_{s+s_1,s_2}(n)
+ z_{s_1 s s_2}(n) + z_{s_1,s+s_2}(n) + z_{s_1 s_2 s}(n)\,.
\label{exam:stuffling}
\end{equation}
Therefore, coefficients of expansion of ratios of products of $P$ functions
can always be made linear in $z$ sums.
In our simple case,
\begin{equation*}
\frac{P_{n-1}(-2\varepsilon)}{P_{n-1}(\varepsilon)} =
\frac{1 - 2 z_1(n) \varepsilon + \cdots}{1 + z_1(n) \varepsilon + \cdots} =
1 - 3 z_1(n) \varepsilon + \cdots
\end{equation*}
Therefore,
\begin{align*}
F' &{}= \sum_{n>0} \frac{1}{n^2}
- 3 \varepsilon \sum_{n>n_1>0} \frac{1}{n^2 n_1}
- 1 - 3 (\zeta_3-1) \varepsilon + \cdots\\
&{}= \zeta_2 - 1 - 3 (\zeta_{21}+\zeta_3-1) \varepsilon + \cdots
\end{align*}

Recalling the duality relation~(\ref{zeta:duality}) $\zeta_{21}=\zeta_3$,
we obtain for our function $F$~(\ref{exam:F})
\begin{align*}
\hspace{-15mm}
F &{}= 2 (2+9\varepsilon+\cdots)
\left[ \zeta_2 - 1 - 3 (2\zeta_3-1) \varepsilon + \cdots \right]\\
&{}= 4 (\zeta_2-1) + 6 (-4\zeta_3+3\zeta_2-1) \varepsilon
+ \cdots
\end{align*}
Thus we have reproduced the first two terms of~(\ref{Gn:Fe}).

\subsection{Expanding hypergeometric functions in $\varepsilon$: algorithm}
\label{S:algo}

Now we shall formulate the algorithm in a general setting.
We want to expand
\begin{equation}
F = \sum_{n=0}^\infty
\frac{\prod_i(m_i+l_i\varepsilon)_n}{\prod_{i'}(m'_{i'}+l'_{i'}\varepsilon)_n}
\label{algo:F}
\end{equation}
in $\varepsilon$.

\textbf{Step 1}. We rewrite all Pochhammer symbols $(m+l\varepsilon)_n$
via $(1+l\varepsilon)_{n+m-1}$:
\begin{equation*}
(m+l\varepsilon)_n =
\frac{(1+l\varepsilon)_{n+m-1}}{(1+l\varepsilon)\cdots(m-1+l\varepsilon)}\,.
\end{equation*}
Introducing the function $P$~(\ref{exam:P}),
we rewrite $F$~(\ref{algo:F}) as
\begin{equation*}
F = \sum_{n=0}^\infty R(n,\varepsilon)
\frac{\prod_i P_{n+m_i-1}(l_i\varepsilon)}{\prod_{i'} P_{n+m'_{i'}-1}(l'_{i'}\varepsilon)}\,,
\end{equation*}
where $R$ is a rational function of $\varepsilon$ and $n$.

\textbf{Step 2}. We expand
\begin{equation*}
R(n,\varepsilon) = R_0(n) + R_1(n) \varepsilon + \cdots
\end{equation*}
and decompose each $R_j(n)$ into partial fractions with respect to $n$.
There is a catch here:
partial-fractioning of a convergent series can split it into a combination of
logarithmically divergent ones.
Therefore, it is necessary to keep the upper limit of summation finite,
and go to the limit only after combining such series together.
Because of shifts of the summation variable, cancellation of divergent series
leaves us with a few terms around the upper limit.
They, however, tend to zero, and this means that we may formally manipulate
logarithmically divergent series as if they were convergent.

Shifting summation indices, we can write $F$~(\ref{algo:F})
as a sum of terms of the form
\begin{equation*}
\sum_{n=n_0}^\infty \frac{1}{n^k}
\frac{\prod_i P_{n+m_i}(l_i\varepsilon)}{\prod_{i'}P_{n+m'_{i'}}(l'_{i'}\varepsilon)}\,.
\end{equation*}

\textbf{Step 3}. We rewrite all $P_{n+m}(l\varepsilon)$ via $P_{n-1}(l\varepsilon)$:
\begin{equation*}
P_{n+m}(l\varepsilon) = P_{n-1}(l\varepsilon)\times
\left(1+\frac{l\varepsilon}{n}\right) \cdots
\left(1+\frac{l\varepsilon}{n+m}\right)\,.
\end{equation*}
Our $F$ becomes a sum of terms of the form
\begin{equation*}
\sum_{n=n_0}^\infty \frac{R(n,\varepsilon)}{n^k}
\frac{\prod_i P_{n-1}(l_i\varepsilon)}{\prod_{i'}P_{n-1}(l'_{i'}\varepsilon)}\,,
\end{equation*}
where $R$ is a rational function of $\varepsilon$ and $n$.
Moving from terms with lower powers of $\varepsilon$
to ones with higher powers, we expand $R$'s in $\varepsilon$:
\begin{equation*}
R(n,\varepsilon) = 1 + R_1(n) \varepsilon + \cdots
\end{equation*}
Terms of higher orders in $\varepsilon$ are decomposed in partial fractions again.
Finally, $F$ becomes a sum of terms of the form
\begin{equation*}
\sum_{n=n_0}^\infty \frac{1}{n^k}
\frac{\prod_i P_{n-1}(l_i\varepsilon)}{\prod_{i'}P_{n-1}(l'_{i'}\varepsilon)}\,.
\end{equation*}

\textbf{Step 4}. Adding and subtracting terms with $n$ from 1 to $n_0-1$,
we rewrite $F$ as a sum of terms of the form
\begin{equation*}
\sum_{n=1}^\infty \frac{1}{n^k}
\frac{\prod_i P_{n-1}(l_i\varepsilon)}{\prod_{i'}P_{n-1}(l'_{i'}\varepsilon)}
\end{equation*}
and rational functions of $\varepsilon$
(which can be trivially expanded in $\varepsilon$).

\textbf{Step 5}. We expand each $P_{n-1}(l\varepsilon)$ as~(\ref{exam:Pe}).
Expansions of ratios of products of $P$ functions contain products
of $z$ sums;
they are reduced to linear combinations of $z$ sums
by stuffling relations (e.g.,~(\ref{exam:stuffling})).
Using
\begin{equation*}
\sum_{n=1}^\infty \frac{1}{n^k} z_{s_1\dots s_j}(n) = \zeta_{k s_1\dots s_j}\,,
\end{equation*}
we can express coefficients of $\varepsilon$ expansion of $F$
as linear combinations of multiple $\zeta$ values
(discussed in Sect.~\ref{S:zeta}) and rational numbers.
\emph{Q.~E.~D.}

A few historical comments.
In 2000, I needed to expand some ${}_3 F_2$ functions of unit argument in $\varepsilon$
(in fact, they were~(\ref{Jn:F}) and~(\ref{In:F})).
I asked David Broadhurst how to do this,
and he replied: just expand Pochhammer symbols in $\varepsilon$,
the coefficients will be expressible via multiple $\zeta$ values.
Following this advise, I implemented the algorithm described above
in \textsf{REDUCE}, and obtained the expansions~(\ref{Jn:F}) and~(\ref{In:F}).
Presentation in Sects.~\ref{S:exam} and~\ref{S:algo}
closely follows my notes from 2000.
Later this algorithm was described as Algorithm A in~\cite{MUW:02}
(this paper also contains more difficult algorithms (B, C, D)
to expand some more complicated sums in $\varepsilon$).
All of these algorithms are implemented in the \textsf{C++} library
\textsf{nestedsums}~\cite{W:02}%
\footnote{Unfortunately, there exist different notations
for multiple $\zeta$ values. We follow~\cite{BBBL:01} here;
in~\cite{MUW:02,W:02} the order of indices is reversed.}.
It is based on the \textsf{C++} computer algebra library \textsf{GiNaC}~\cite{BFK:02}.
In order to use \textsf{nestedsums} with recent versions of \textsf{GiNaC},
you should download a recent version from the URL shown in~\cite{W:02}.

Unfortunately, no general algorithm is known to expand in $\varepsilon$
hypergeometric functions of unit argument some of whose indices tend to
half-integers at $\varepsilon\to0$.
Such functions appear in on-shell propagator calculations,
see Sects.~\ref{S:M2}, \ref{S:M2m}.
Known expansions show that many constants more complicated than
multiple $\zeta$ values appear.

\FloatBarrier
\section*{Acknowledgements}
\addcontentsline{toc}{section}{Acknowledgements}

I am grateful to V.M.~Braun for providing details
of the calculation~\cite{BB:94} of $I(1,1,1,1,n)$;
to D.J.~Broadhurst, K.G.~Chetyrkin, A.~Czarnecki, A.I.~Davydychev
for collaboration on various multiloop projects;
to A.V.~Kotikov for the discussion of~\cite{K:96};
to the organizers of Calc-03 in Dubna for inviting me
to give these lectures and for organizing this excellent school.

\FloatBarrier

\end{document}